\documentclass{CSML}

\def\dOi{4(4:9)2008}
\lmcsheading%
{\dOi}
{1--27}
{}
{}
{Oct.~30, 2007}
{Apr.~19, 2016}
{}

\ACMCCS{[{\bf Theory of computation}]: Models of computation---Quantum
  computation theory; Semantics and reasoning---Program
  semantics---Categorical semantics; [{\bf Hardware}]: Emerging
  technologies---Quantum technologies---Quantum computation---Quantum
  communication and cryptography; [{\bf Security and privacy}]:
  Cryptography---Information-theoretic techniques}

\usepackage{xypic,amssymb,pifont,color,hyperref}
\newcommand{\blue}[1]{{\color{black}{#1}}}
\newcommand{\one}{\ensuremath{\text{\ding{192}}}}
\newcommand{\two}{\ensuremath{\text{\ding{193}}}}
\newcommand{\three}{\ensuremath{\text{\ding{194}}}}
\newcommand{\four}{\ensuremath{\text{\ding{195}}}}
\newcommand{\five}{\ensuremath{\text{\ding{196}}}}
\newcommand{\six}{\ensuremath{\text{\ding{197}}}}
\newcommand{\seven}{\ensuremath{\text{\ding{198}}}}
\newcommand{\after}{\circ}
\newcommand{\cat}[1]{\ensuremath{\mathbf{#1}}}
\newcommand{\Cat}[1]{\ensuremath{\mathrm{\textbf{#1}}}}
\newcommand{\Set}{\Cat{Set}}
\newcommand{\idmap}[1][]{\ensuremath{\mathrm{id}_{#1}}}
\newcommand{\id}[1]{\ensuremath{\mathrm{id}}}
\newcommand{\op}{\ensuremath{^{\mathrm{op}}}}
\newcommand{\field}[1]{\ensuremath{\mathbb{#1}}}
\newcommand{\inprod}[2]{\ensuremath{\langle #1\,|\,#2 \rangle}}
\newcommand{\powerset}{\ensuremath{\mathcal{P}}}
\newcommand{\name}[1]{\ensuremath{\ulcorner #1 \urcorner}}
\newcommand{\coname}[1]{\ensuremath{\llcorner #1 \lrcorner}}
\newcommand{\tensor}{\ensuremath{\otimes}}
\newcommand{\norm}[2][]{\ensuremath{\|#2\|_{#1}}}
\newcommand{\colim}{\ensuremath{\mathop{\mathrm{colim}}}}
\newcommand{\ie}{\textit{i.e.}~}
\newcommand{\eg}{\textit{e.g.}~}

\newcommand{\cpt}{\ensuremath{_{\mathrm{cpt}}}}
\newcommand{\spn}{\ensuremath{\mathrm{span}}}

\begin{document}

\title[Compactly Accessible Categories and Quantum Key Distribution]
      {Compactly Accessible Categories\\and Quantum Key Distribution}

\author[Chris Heunen]{Chris Heunen}
\address{Institute for Computing and Information Sciences\\
         Radboud University, Nijmegen, the Netherlands}


\begin{revision}
  This is a revised an corrected version of the article originally
  published on November 17, 2008.
\end{revision}

\begin{abstract}
  \noindent Compact categories have lately seen renewed interest via 
  applications to quantum physics. Being essentially
  finite-dimensional, they cannot accomodate (co)limit-based
  constructions. For example, they cannot capture protocols such as
  quantum key distribution, that rely on the law of large numbers. To
  overcome this limitation, we introduce the notion of a compactly
  accessible category, relying on the extra structure of a
  factorisation system. This notion allows for infinite dimension while
  retaining key properties of compact categories: the main technical
  result is that the choice-of-duals functor on the compact part extends
  canonically to the whole compactly accessible category. As an
  example, we model a quantum key distribution protocol and prove its
  correctness categorically.
\end{abstract}

\maketitle

\section{Introduction}

Compact categories were first introduced in 1972 as a class of
examples in the context of the coherence problem~\cite{kelly}. They
were subsequently studied first categorically~\cite{day,kellylaplaza},
and later in relation to linear logic~\cite{seely}. Interest has
rejuvenated  since the exhibition of another aspect: compact
categories provide a semantics for quantum
computation~\cite{abramskycoecke,selinger}.  The main virtue of
compact categories as models of quantum computation is that from very
few axioms, surprisingly many consequences ensue that were postulates
explicitly in the traditional Hilbert space formalism, \eg
scalars~\cite{abramsky}. Moreover, the connection to linear logic 
provides quantum computation with a resource sensitive type theory of
its own~\cite{duncan}.  

Much of the structure of compact categories is due to a seemingly
ingrained `finite-dimensionality'. This feature is most apparent in
the prime example, the category of vector spaces
and linear maps. As we will see, the only compact objects in this
category are the finite-dimensional vector spaces. This poses no
problems when applied to quantum computation, where the amount of
memory is physically bounded anyway. However, the employment of compact
categories is sometimes optimistically publicised as providing
`a semantics for quantum protocols', or even `axiomatics for
quantum physics'. For these general purposes, a fixed finite dimension
is a severe limitation since it rules out (co)limit constructions and
arguments. In fact, the simplest possible physical 
situation, that of a single free-moving particle in three-dimensional
space, is already modeled by the infinite-dimensional space
$L^2(\field{R}^3)$ of observables in traditional quantum
physics~\cite{vonneumann}. Likewise, an important class of quantum
protocols relies on the law of large numbers. They utilise the
probabilistic nature of quantum physics to ensure that their goal is
reached after sufficiently many tries. In fact, the two most-cited
papers in quantum cryptography to date, describing quantum key
distribution protocols, are of this kind~\cite{bennettbrassard,ekert}. 

There have been earlier attempts to remedy the above
limitation. Although he did not have the quantum setting in mind, Barr
gave a construction to embed a category with certain minimal
properties fully into a complete and cocomplete category that is
*-autonomous, a notion closely related to
compactness~\cite{barr}. However, as we will see, the important
category of Hilbert spaces and bounded maps, that is the traditional
model of quantum physics, is neither complete nor cocomplete.

Another proposal revolves around the use of nuclear
ideals~\cite{abramskyblutepanangaden,blute}. Analogous to ring theory, an
ideal in this setting is a set of morphisms that is closed under
composition with arbitrary morphisms. The adjective nuclear means that
the key property that enables compact categories to model quantum
protocols is postulated to hold for all morphisms in the ideal. 
This seems to be the right environment to study properties of morphisms in a
quantum setting. For example, a very natural characterisation of
trace-class morphisms emerges. However, it also forces one to consider
two layers, the category and the nuclear ideal, and possible coherence
with the ideal is a distraction when working with notions that are
more naturally defined on the category. For example, any bounded map
between Hilbert spaces has a dual map (in the opposite direction
between the dual spaces), not just the Hilbert-Schmidt maps (that form
a nuclear ideal).

The present work introduces the notion of a compactly accessible
category in order to overcome the above limitation. It retains certain
key properties of compact categories, and simultaneously allows for 
infinite dimension.   
The main idea is to relax the requirement that every object is compact
to the requirement that every object is a directed colimit of compact
ones, imitating the fact that every vector space is the directed
colimit of its finite-dimensional subspaces. Categories in which every
object is a directed colimit of finitely presentable ones are
well-known as accessible categories, and a polished theory has 
developed around them~\cite{gabrielulmer,adamekrosicky}. We
weaken the concept of finitely presentable object to that of a
compactly presentable one, to ensure that the key properties of
compact categories are inherited by compactly accessible categories.
The central novel ingredient is the extra structure of a factorisation
system. This approach provides a proper category in which to model quantum
protocols, and hence is automatically compositional --- as opposed to
ideals that typically do not include all identity
maps~\cite{blutepanangadenpronk}. 
Physically, directed colimits provide the intuition of `time'.
The main result, that justifies our definition of compactly accessible
category, is Theorem~\ref{thm:starfunctor}. It shows that the
choice-of-duals functor on the compact part extends canonically to the
whole compactly accessible category. It is remarkable that 
this canonical extension of the choice-of-duals functor in the
category of Hilbert spaces with its canonical factorisation system in
fact provides an equivalence with the opposite category. This
is another indication that the axiomatic structure of compactly
accessible categories is on target.
Moreover, Theorem~\ref{thm:daggercommuteswithstar} proves that 
if the choice-of-duals functor commutes with a dagger functor on the
compact part, then so does its canonical extension. The latter is
important for the modeling of quantum physics.
As an example, we model a quantum
key distribution protocol and prove its correctness categorically. 

Section~\ref{sec:quantumkeydistribution} first introduces the
quantum key distribution protocol mentioned
above. We recall the necessary details of compact categories in
Section~\ref{sec:compactobjectsandcompactcategories}.
Subsequently, Section~\ref{sec:compactlypresentableobjects}
builds up to the notion of compactly presentable object, and
Section~\ref{sec:compactlyaccessiblecategories} then defines compactly
accessible categories and explores their structure. Dagger structure
is added in Section~\ref{sec:daggercompactlyaccessiblecategories}, and
Section~\ref{sec:quantumkeydistributioncategorically} models
and proves correct the protocol described in
Section~\ref{sec:quantumkeydistribution}. Finally,
Section~\ref{sec:conclusion} concludes.

\section{Quantum key distribution}
\label{sec:quantumkeydistribution}

\begin{figure}
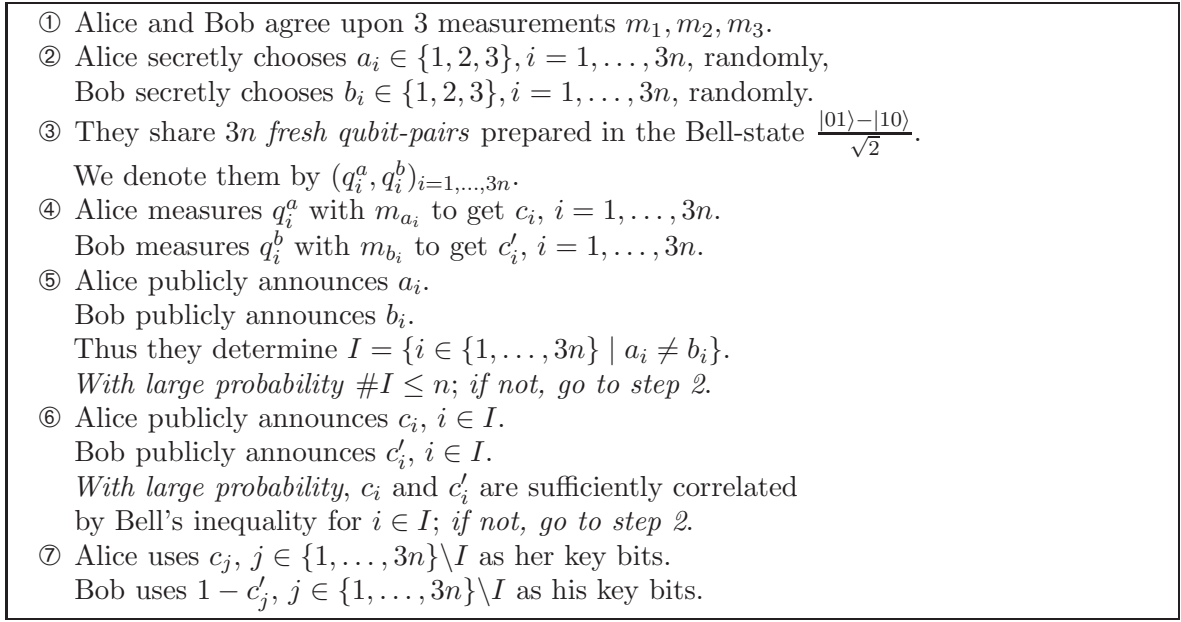

  \centering
  \framebox{
  \begin{minipage}{1.0\linewidth}
   \begin{enumerate}
    \item[\one] Alice and Bob agree upon 3 measurements $m_1, m_2, m_3$.
    \item[\two] Alice secretly chooses $a_i \in \{1,2,3\},
      i=1,\ldots,3n$, randomly, 
      \\ Bob secretly chooses $b_i \in \{1,2,3\}, i=1,\ldots,3n$, randomly.
    \item[\three] They share $3n$ \emph{fresh qubit-pairs} prepared in the
      Bell-state $\frac{|01\rangle-|10\rangle}{\sqrt{2}}$. \\
      We denote them by $(q_i^a, q_i^b)_{i=1,\ldots,3n}$.
    \item[\four] Alice measures $q_i^a$ with $m_{a_i}$ to get $c_i$,
      $i=1,\ldots,3n$. \\ Bob measures $q_i^b$ with $m_{b_i}$ to get
      $c_i'$, $i=1,\ldots,3n$.
    \item[\five] Alice publicly announces $a_i$. \\ Bob publicly announces
      $b_i$. \\ Thus they determine $I=\{i \in \{1,\ldots,3n\} \mid
      a_i \neq b_i\}$. \\ \emph{With large probability} $\#I \leq n$;
      \emph{if not, go to step 2}. 
    \item[\six] Alice publicly announces $c_i$, $i \in I$. \\ Bob publicly
      announces $c_i'$, $i \in I$. \\ \emph{With large probability}, $c_i$ and
      $c_i'$ are sufficiently correlated \\ by Bell's
      inequality for $i \in I$; \emph{if not, go to step 2}. 
    \item[\seven] Alice uses $c_j$, $j \in \{1,\ldots,3n\}\backslash I$ as her
      key bits. \\ 
      Bob uses $1-c_j'$, $j \in \{1,\ldots,3n\}\backslash I$ as his key bits. 
   \end{enumerate}
  \end{minipage}}
  \caption{A quantum protocol to obtain a $2n$-bits shared secret
           key~\cite{ekert}.} 
  \label{fig:e91}
\end{figure}

\emph{Quantum key distribution} is the name for a collection of protocols
that provide two parties using a quantum channel between them with a
shared binary string, unknowable to anyone else. Moreover, such a
scheme must be proven inherently secure by the laws of
nature, \ie not depending on any unsolved or computationally
unfeasible mathematical problems. The most well-known protocol in this
family is that of Bennett and Brassard~\cite{bennettbrassard}, which
essentially relies on Bell's inequality and the law of large numbers
to provide secure keys.\footnote{Such a protocol, like
Diffie-Hellmann's~\cite{diffiehellman}, regulates key distribution,
but gives no guarantee about authenticity of the two parties involved.} 
There are several improvements upon this protocol. Especially
Ekert~\cite{ekert} developed a very nice simplification, which is
outlined in Figure~\ref{fig:e91}. As Bell's inequality provides a
means to verify that two qubits are `correlated enough', eavesdroppers
can be detected with large probability. The law of large numbers thus
ensures that this protocol works (up to a negligable probability that
can be specified in advance). Notice that because of the possible
jump back in steps $\five$ and $\six$, the number of fresh qubit-pairs
needed is not known in advance. 

The protocol in Figure~\ref{fig:e91} will be used in
Section~\ref{sec:quantumkeydistributioncategorically} as an example 
that can be modeled by compactly accessible categories. As such, we
need to distinguish between correctness and security. A quantum key
distribution protocol is correct if both parties end up with the same
key in every run, \ie when $c_j=1-c_j'$ for all $j \in
\{1,\ldots,3n\}\backslash I$ and every choice of $m_i$, $a_i$ and
$b_i$ in Figure~\ref{fig:e91}. It is secure 
when a potential eavesdropper cannot learn any of the key bits. In
this instance, the security relies on Bell's inequality. Thus in this case
one could say that correctness is a qualitative notion, and security a
quantitative one. Because the entire purpose of the categorical
approach is to abstract away from quantitative details like scalar
factors, we will focus on correctness, and forget about the classical
calculation in step $\six$. Since the center of attention in 
this article is the elimination of finite-dimensionality, we will also
not concern ourselves too much with the classical communication that
is most noticable in step $\five$. The point is just to show that
compactly accessible categories are able to model protocols that need
an a priori unknown number of resources.

\section{Compact objects and compact categories}
\label{sec:compactobjectsandcompactcategories}

This section recalls the concept of a compact category, by considering
the required properties separately per object. It also reviews the key
features of compact categories that are so important to model quantum
protocols. 

\begin{defi} 
\label{def:compactobject}
  An object $X$ of a symmetric monoidal category \cat{C} is said to be
  \emph{compact} when there are an object $Y \in \cat{C}$ and morphisms
  $\eta:I \to Y \tensor X$ and $\varepsilon:X \tensor Y \to I$
  such that the following diagrams commute.
  \begin{equation}
  \label{eq:compactobject}\raise5ex\hbox{\xymatrix@C-3.5ex{
       X \ar^-{\cong}[r] \ar@{=}[d]
     & X \tensor I \ar^-{\idmap \tensor \eta}[rr]
    && X \tensor (Y \tensor X) \ar^-{\cong}[d] 
    && Y \ar^-{\cong}[r] \ar@{=}[d]
     & I \tensor Y \ar^-{\eta \tensor \idmap}[rr]
    && (Y \tensor X) \tensor Y \ar^-{\cong}[d] \\
       X 
     & I \tensor X \ar^-{\cong}[l]
    && (X \tensor Y) \tensor X \ar^-{\varepsilon \tensor \idmap}[ll]
    && Y
     & Y \tensor I \ar^-{\cong}[l]
    && Y \tensor (X \tensor Y) \ar^-{\idmap \tensor \eta}[ll]
  }}\end{equation}
  A symmetric monoidal category is called compact when all its objects
  are.
\end{defi}

For a given compact object $X$, the object $Y$ of the previous
definition is called a \emph{dual object} for $X$. Such dual objects
are unique up to isomorphism~\cite[Proposition~2.7]{duncan}. A chosen
dual object for $X$ is usually denoted by $X^*$. 
Notice that $I$ is a compact object in any strict symmetric monoidal
category, with $I^*=I$. Also, if $X$ is compact, then so is
$X^*$. Moreover, any compact object $X$ is isomorphic to its double
dual $X^{**}$~\cite[Proposition~2.13]{duncan}. Let us see what the
compact objects (and their dual objects) are in a few example categories.

\begin{exa}
\label{ex:compactobject:posetalgroup}
  In a posetal symmetric monoidal category,
  diagrams~\eqref{eq:compactobject} say that an object $X$ is compact
  precisely when there is an object $X^*$ such that $X^* \tensor X = I
  = X \tensor X^*$.    
  Any ordered commutative monoid is such a category, where the order
  induces the morphisms, and the monoid multiplication and
  unit provide symmetric monoidal structure. 
  Hence, the compact objects in an ordered commutative monoid, seen as
  a posetal category, are precisely its invertible elements. 
  Thus any ordered Abelian group induces a compact category; an
  Abelian group is partially ordered if and only if it is
  torsion-free~\cite[Theorem 1.1.3]{murarhemtulla}. This example has
  been studied more generally under the name `Lambek
  pregroups'~\cite{sadrzadeh}.  
\end{exa}

\begin{exa} 
\label{ex:compactobject:Rel}
  Denote by $\Cat{Rel}$ the category with sets for objects, and
  relations $R \subseteq X \times Y$ for morphisms $X \to
  Y$. The composition of $\xymatrix@1{X \ar^-{R}[r] & Y \ar^-{S}[r] & Z}$
  is the usual relational composition
  \[
    S \after R = \{ (x,z) \in X \times Z \mid \exists_{y \in Y}. (x,y)
    \in R \wedge (y,z) \in Z \}.
  \]
  This category is symmetric monoidal by the usual set-theoretic
  product, with the singleton set $\{*\}$ as its neutral
  element. Every object $X$ in $\Cat{Rel}$ is compact: by defining
  $X^*=X$ and
  \begin{align*}
    \eta_X & = \{(*,(x,x)) \;\colon x \in X\}, \\
    \varepsilon_X & = \{((x,x),*) \;\colon x \in X\},
  \end{align*}
  one easily verifies that diagrams~\eqref{eq:compactobject}
  commute. Hence $\Cat{Rel}$ is a compact category. 

  This example can be generalized to the Kleisli category of the monad
  on $\Set$ given by $\powerset(M \times -)$ for an arbitrary
  commutative monoid $M$ instead of the trivial monoid. It can also be
  generalized to the category of relations on an arbitrary regular
  category~\cite{carbonikasangianstreet}. In both generalized
  categories, every object is compact.
\end{exa}

At first sight, one might expect that the category $\Cat{Sup}$ of
complete lattices and sup-preserving functions is compact, but it is
not~\cite[page 99]{barr}. Its largest compact subcategory is that of
complete atomic boolean lattices and sup-preserving functions; this
category is equivalent to $\Cat{Rel}$.

\begin{exa} 
\label{ex:compactobject:Vect}
  Denote by $\Cat{Vect}$ the category of complex vector spaces and
  linear maps. It is a symmetric monoidal category by the usual tensor
  product of vector spaces, with the complex field $\field{C}$ as
  unit. Any finite-dimensional vector space $X$ is a compact
  object in this category as follows. Let $X^*$ be the dual vector
  space $\{f:X \to \field{C} \mid f \mbox{ linear}\}$. If $(e_i)$ is a
  basis for $X$, then the functionals $\overline{e}_i$ determined by
  $\overline{e}_i(e_j)=\delta_{ij}$ form a basis for $X^*$. Define
  $\eta_X$ and $\varepsilon_X$ by linear extension of
  \begin{align*}
    \eta_X(1) & = \sum_{i=1}^{\dim(X)} \overline{e}_i \tensor e_i, \\
    \varepsilon_X(e_i \tensor \overline{e}_j) & = \overline{e}_j(e_i).
  \end{align*}
  Diagrams~\eqref{eq:compactobject} are readily seen to commute.

  However, an infinite-dimensional vector space cannot be isomorphic
  to its double dual because of a well-known cardinality
  argument~\cite[Theorem IX.2]{jacobson} that we sketch briefly. Let
  $X$ be an infinite-dimensional vector space, and choose a basis $B$
  for it. Then $X \cong \coprod_B \field{C}$, and so $X^* \cong \prod_B
  \field{C}$~\cite[Proposition 20.2]{andersonfuller}. So $\dim(X)
  \lneqq \dim(X^*) \lneqq \dim(X^{**})$, whence $X \not\cong X^{**}$ and
  $X$ is not a compact object in $\Cat{Vect}$.\footnote{For
  completeness' sake, let us recall that even for a finite-dimensional 
  vector space $X$, the isomorphism $X \cong X^*$ is not natural,
  although $X \cong X^{**}$ is~\cite[Section VII.4]{maclane}.}

  Hence the full subcategory $\Cat{fdVect}$ of $\Cat{Vect}$ containing
  only the finite-dimensional vector spaces is the largest compact
  subcategory of $\Cat{Vect}$.

  This example can be generalized to the category of projective
  modules over a given semiring: the compact objects in that
  category are precisely the finitely generated ones.
\end{exa}

\begin{exa} 
\label{ex:compactobject:Hilb}
  As an extension of the previous example, consider the category
  $\Cat{Hilb}$ of Hilbert spaces. Its morphisms are the bounded linear
  maps, \ie linear functions $f:X \to Y$ for which there is
  a constant $\norm{f}$ such that $\norm{f(x)} \leq
  \norm{f}\norm{x}$ 
  for all $x \in X$. It is a symmetric monoidal category with the usual tensor
  product and the complex field $\field{C}$ as unit.
  Any finite-dimensional Hilbert space $X$ is a compact object in
  this category as follows. Let $X^*$ be the conjugate of the dual
  space $\{f:X \to \field{C} \mid f \mbox{ bounded linear}\}$, \ie it
  has the same additive group as the dual space, but conjugated scalar
  multiplication. Then $X^* \tensor X$ is isomorphic to the Hilbert
  space of all Hilbert-Schmidt maps $X \to
  X$~\cite{kadisonringrose}. Define $\eta_X$ by letting 1 correspond
  to the identity map under this isomorphism and extending linearly
  and continuously, and define $\varepsilon_X$ as the adjoint of
  $\eta_X$. Then diagrams~\eqref{eq:compactobject} commute. Since the
  identity map on $X$ is a Hilbert-Schmidt map if and only if $X$ is
  finite-dimensional, this recipe for obtaining compact structure on
  $X$ only works for finite-dimensional $X$. In other words,
  $\Cat{fdHilb}$ is a compact full subcategory of $\Cat{Hilb}$. 
  Moreover, as Proposition~\ref{prop:propertiesofcompactobjects}(d)
  below shows, a compact full subcategory of $\Cat{Hilb}$ is
  necessarily closed. Since only the Hilbert-Schmidt functions form
  a Hilbert space again~\cite{kadisonringrose}, a compact full
  subcategory of $\Cat{Hilb}$ must consist of objects between which
  all continuous linear functions are automatically
  Hilbert-Schmidt. That is, the largest compact full subcategory of
  $\Cat{Hilb}$ is $\Cat{fdHilb}$.

  This example can be generalized to the category of unitary
  representations of a given topological group: the compact objects in
  that category are precisely the 
  representations with a finite-dimensional target space.\footnote{In
  fact, this is where the compactness terminology seems to have
  originated: the group $G$ can be reconstructed from the described
  category $\Cat{fdURep}(G)$ when it is
  compact~\cite{doplicherroberts,mueger}. Hence the name transferred 
  from the group to categories resembling $\Cat{fdURep}(G)$.
  Alternatively, one could observe that being a Hausdorff space, a
  Hilbert space's unit ball is compact if and only if it is
  finite-dimensional. Finally, a Hilbert space is locally compact if and
  only if it is finite-dimensional~\cite[Problem~10]{halmos:problembook}.}
\end{exa}

Introducing the notation $\cat{C}\cpt$ for the full subcategory of
compact objects of $\cat{C}$, the previous examples thus show that
$\Cat{Rel}\cpt=\Cat{Rel}$, $\Cat{Vect}\cpt=\Cat{fdVect}$, and
$\Cat{Hilb}\cpt=\Cat{fdHilb}$.
This relates to order theory, in which `finite element' and `compact
element' are used interchangeably~\cite{johnstone,abramskyjung}.
On the one hand the name `finite object' or `finite-dimensional
object' would also be apt in our case, but on the other hand it would
be confusing since a compact object in $\Cat{Rel}$ can be infinite as
a set. 

As an example of the properties of compact objects, we mention the
following. They are standard results; here we formulate them for
compact objects (instead of for compact categories).

\begin{prop} 
\label{prop:propertiesofcompactobjects}
  \cite{lindner2}
  Let $\cat{C}$ be a symmetric monoidal category.
  \begin{enumerate}[label=\({\alph*}]
    \item[(a)] If $X \in \cat{C}\cpt$, then $\cat{C}(X,Y) \cong
      \cat{C}(I,X^* \tensor Y)$ for all $Y \in \cat{C}$.
    \item[(b)] If $Y \in \cat{C}\cpt$, then $\cat{C}(X,Y) \cong
      \cat{C}(X \tensor Y^*, I)$ for all $X \in \cat{C}$.
    \item[(c)] An object $X \in \cat{C}$ is compact if and only if there is
      an $Y \in \cat{C}$ such that\\ $\cat{C}(X \tensor Y, I) \cong
      \cat{C}(X,X) \cong \cat{C}(I,Y \tensor X)$.
    \item[(d)] An object $X \in \cat{C}$ is compact iff
      there is a $Y \in \cat{C}$ such that $X \tensor (-)$ is left
      adjoint to $Y \tensor (-)$. In that case, $X \tensor (-)$ is
      also right adjoint to $Y \tensor (-)$.
    \item[(e)] If $X \in \cat{C}\cpt$, then $(-) \tensor X:\cat{C} \to
      \cat{C}$ is both continuous and cocontinuous.
  \qed
  \end{enumerate}
\end{prop}

The crucial property of a compact category is that a choice of dual
objects $X^*$ extends functorially, as follows.

\begin{prop}
\label{prop:dualmorphismcpttocpt}
  \cite{kellylaplaza}
  For a morphism $f:X \to Y$ between compact objects $X,Y$ in some
  category \cat{C}, define $f^*:Y^* \to X^*$ as the composite
  \[\xymatrix{
      Y^* \cong Y^* \tensor I \ar^-{\idmap \tensor \eta_X}[r]
    & Y^* \tensor (X \tensor X^*) 
      \ar^-{\idmap \tensor f \tensor \idmap}[r] 
    & (Y^* \tensor Y) \tensor X^* \ar^-{\varepsilon_Y \tensor
    \idmap}[r]
    & I \tensor X^* \cong X^*.
  }\]
  This defines a functor $(-)^*:\cat{C}\cpt\op\to\cat{C}\cpt$.
  \qed
\end{prop}

For future reference, let us mention that the correspondence in
Proposition~\ref{prop:propertiesofcompactobjects}(a,b) of morphisms $f:X
\to Y$ to their \emph{name}s $\name{f}:I \to X^* \tensor Y$ and to
their \emph{coname}s $\coname{f}:X \tensor Y^* \to I$ 
satisfies~\cite[Lemma 2.18]{duncan}:
\begin{equation}
  \label{eq:absorption} 
    (\idmap \tensor g) \after \name{f} 
  = \name{g \after f} 
  = (f^* \tensor \idmap) \after \name{g}.
\end{equation}
Moreover, the choice-of-duals functor $(-)^*$ preserves limits and
colimits. When $D:\cat{J} \to \cat{C}$ is any
diagram in a compact category, we can speak of its dual diagram
$D^*:\cat{J} \to \cat{C}\op$ determined by $D^*=(-)^* \after D$. 
This construction extends to \emph{compact diagrams} in any category:
we say a diagram in any category \cat{C} is compact if it factors
through $\cat{C}\cpt$. We use the term \emph{compact (co)limit} for a
(co)limit of a compact diagram.

\begin{prop}
\label{prop:compactcategory:starisequivalence}
  If \cat{C} is a compact category, $(-)^*:\cat{C}\op\to\cat{C}$
  preserves limits and colimits, \ie for any diagram
  $D:\cat{J}\to\cat{C}$, we have $(\lim(D))^* \cong \lim(D^*)$ and
  $(\colim(D))^* \cong \colim(D^*)$ in $\cat{C}\op$.
\end{prop}
\proof
  If $\cat{C}$ is a compact category, the functor $(-)^* \colon
  \cat{C}\op \to \cat{C}$ is an equivalence of
  categories~\cite[Proposition 2.13]{duncan}. 
\qed

\section{Compactly presentable objects}
\label{sec:compactlypresentableobjects}

The next section discusses a kind of category in which every object is
a co\-limit of compact ones. However, taking colimits of just compact
objects is not enough to retain a `choice-of-duals-functor' as in
Proposition~\ref{prop:dualmorphismcpttocpt}. 
This section strengthens the notion of compact object as the
constituent of the colimits accordingly, drawing
inspiration from the notion of finitely presentable object. An extra
ingredient is the structure of a factorisation system.

\subsection{Finitely presentable objects}

Intuitively, a finitely presentable object is one that can be
described algebraically using a finite number of generators and
finitely many equations~\cite{adamekrosicky}. Recall that a 
preorder is \emph{directed} when every two elements have a
common upper bound; a \emph{directed colimit} is a colimit of a
directed preorder considered as a diagram. Likewise, a preorder
is codirected when its opposite is directed, and a codirected limit is
a limit of a codirected preorder considered as a diagram.

\blue{
\begin{defi}
  A \emph{normed category} is a category $\cat{C}$ together with an ordered monoid $R$ and functions $\|-\| \colon \cat{C}(X,Y) \to R$ satisfying $\| \idmap[X] \|=1$ and $\| g \circ f \| \leq \|g\| \cdot \|f\|$ for composable morphisms $f,g$.
  A \emph{normed monoidal category} additionally satisfies $\|f \otimes g\|=\|f\| \cdot \|g\|$ for all morphisms $f,g$.
  Finally, a \emph{compact monoidal category} additionally satisfies $\|f^*\|=\|f\|$.
\end{defi}

A \emph{bounded diagram} in a normed category is a diagram such that there exists a bound $r \in R$ with $\|f\| \leq r$ for all morphisms $f$ in the diagram.
A (co)cone $(d_i)$ on a diagram is bounded when there exists a bound $r \in R$ such that $\|d_i\| \leq r$ for all $i$. A \emph{bounded (co)limit} is a universal bounded (co)cone.
Any category is normed in $(\{0,1\},\cdot,\leq)$ with $\|f\|=1$ for any morphisms $f$, in which case bounded (co)limits are precisely ordinary (co)limits.
}

\begin{defi}
  An object $X$ in a \blue{normed} category $\cat{C}$ is called \emph{finitely
  presentable} when the hom-functor\footnote{If needed, one should
  replace $\Cat{Set}$ by some suitably larger universe.}
  $\cat{C}(X,-):\cat{C} \to \Cat{Set}$ preserves \blue{bounded} directed
  colimits.\footnote{This definition can be extended to
  $\lambda$-presentable, for a regular cardinal $\lambda$. Finite
  presentability then coincides with $\omega$-presentability. Later
  notions, like finitely accessible category, can also be extended,
  but for the sake of clarity of presentation we do not do so in this
  article.}   
\end{defi}

Writing this out, we see that $X$ is finitely presentable when for any
\blue{bounded} directed diagram $D:\cat{J} \to \cat{C}$, any \blue{bounded} colimit cocone $d_j:D(j)
\to Y$ and any morphism $f:X \to Y$, there are $j \in \cat{J}$ and a
morphism $g:X \to D(j)$ such that $f = d_j \after g$. Moreover, this
morphism $g$ is essentially unique, in the sense that if $f=d_j \after
g = d_j \after g'$, then $D(j \to j') \after g = D(j \to j') \after
g'$ for some $j' \in \cat{J}$. 
\[\xymatrix{
    D(1) \ar[r] \ar_-{d_1}[dr] 
  & D(2) \ar[r] \ar_-{d_2}[d] 
  & D(3) \ar[r] \ar_-{d_3}[dl]
  & \cdots \ar[r] 
  & D(j) \ar[r] \ar_-{d_j}[dlll]
  & \cdots \\
  & Y & & & X \ar^-{f}[lll] \ar@{-->}_-{g}[u]
}\]
The following example shows that finite presentability is certainly an 
interesting property in the context of compact objects in categories.

\begin{exa}
\label{ex:finitelypresentableobject:posetalgroup}
  In the posetal category induced by an ordered commutative monoid as in
  Example~\ref{ex:compactobject:posetalgroup}, an object $X$ is finitely
  presentable precisely when in case $X$ is smaller than a supremum of
  some directed set $D$, it is already smaller than
  some element of $D$.  
  (This is closely related to a compact or finite element of a lattice
  in the order theoretical sense~\cite{johnstone,abramskyjung}.)
\end{exa}

The next example shows that in some categories, the finitely
presentable objects are precisely the compact ones.

\begin{exa} 
\label{ex:finitelypresentableobject:SetVectHilb}
  An object in $\Cat{Set}$ is finitely presentable if and only if it
  is a finite set. An object in $\Cat{Vect}$ or $\Cat{Hilb}$\blue{, with the $\mathbb{R}$-valued operator norm,} is
  finitely presentable if and only if it is finite-dimensional.
\end{exa}

However, the following example shows that $\Cat{Rel}$ has only one
finitely presentable object. This contrasts sharply with
Example~\ref{ex:compactobject:Rel}, that shows that every object in
$\Cat{Rel}$ is compact.

\begin{exa} 
\label{ex:finitelypresentableobject:Rel}
  The only finitely presentable object of \Cat{Rel} is the empty set.
\end{exa}
\proof
  Since $\emptyset$ is an initial object in \Cat{Rel}, any morphism
  $\emptyset \to \colim(D)$ is the empty relation, which factors
  uniquely through any $D(j)$.

  Conversely, suppose that $X$ has an element $x$. Consider the
  directed diagram $D:\field{N} \to \Cat{Rel}$, where $\field{N}$ is a
  partial order seen as a category, determined by
  $D(n)=\{0,\ldots,n-1\}$ and $D(n \to m) = \{ (i,i) \colon
  i=0,\ldots,n-1 \}$. Its colimit in \Cat{Rel} is
  $\field{N}$, with colimit cocone $d_n = \{ (i,i)
  \colon i=0,\ldots,n-1 \} \subseteq D(i) \times \field{N}$. Define a
  relation $R \subseteq X \times \field{N}$ by $R = \{ (x,n) \colon n
  \in \field{N} \}$. If this relation were to factor through any $D_n$
  then its image would have to be finite, which it is not. Hence $X$
  is not finitely presentable.
\qed

It is interesting to note that, in a sense, the notion of compact
object is stronger than that of finitely presentable object.
By Proposition~\ref{prop:propertiesofcompactobjects}(d), the compact
part $\cat{C}\cpt$ of any category $\cat{C}$ is monoidal closed, and
hence enriched over itself. 
Since the $\cat{C}$-functor $\cat{C}(X,-)$ is $\cat{C}$-cocontinuous
for $X \in \cat{C}\cpt$~\cite[Proposition 6]{lindner2}, one 
might think that Proposition~\ref{prop:propertiesofcompactobjects}(e)
implies that every compact object of a category is finitely
presentable. However, there is a distinction between cocontinuity of
$\cat{C}(X,-)$ in this enriched setting~\cite{lindner} and `ordinary'
finite presentability. For example, sets and
relations can be seen as an ordinary $\Cat{Set}$-category $\Cat{Rel}$
with hom-sets $\powerset(X \times Y)$, but also as a
$\Cat{Rel}$-category $\underline{\Cat{Rel}}$ with hom-objects $X
\times Y$. However, cocontinuity in $\Cat{Rel}$ is different entirely
from cocontinuity in  $\underline{\Cat{Rel}}$: the former just means
that $X \times (-)$ preserves all colimits in $\Cat{Rel}$, whereas the
latter means that $\Cat{Rel}(X,-)$ preserves all colimits in
$\Cat{Rel}$. In other words:
\begin{align*}
  X \times \colim(D) & \cong \colim X \times D(j), \mbox{ but} \\
  \powerset(X \times \colim(D)) & \not\cong \colim \powerset(X \times
  D(j)), \mbox{ except for }X=\emptyset, 
\end{align*}
where $D:\cat{J}\to\cat{Rel}$ is a diagram, and the
colimit is taken in $\Cat{Rel}$. This explains why $\Cat{Rel}$ has
only one finitely presentable object and every object is compact, but
every object of $\underline{\Cat{Rel}}$ is compact and finitely
presentable.

\subsection{Factorisation systems}

To arrive at a suitable notion that is stronger than compactness of
objects but retains the essential properties of finite presentability,
we recall a concept that was popularised by Freyd and
Kelly~\cite{freydkelly} but whose origins can be traced back to
Mac Lane~\cite{maclane:duality} and Isbell~\cite{isbell}
(see also~\cite[Exercises~5.5]{barrwells} or~\cite[Section~5.5]{borceux}).

\begin{defi} 
  A \emph{weak factorisation system} $(E,M)$ for a category \cat{C}
  consists of two classes of morphisms $E$ and $M$ of \cat{C} such that
  \begin{itemize}
    \item $E$ and $M$ both contain all isomorphisms of \cat{C}, and
      are closed under composition;
    \item Every morphism $f$ of \cat{C} can be factored as $f=m \after
      e$ for some $m \in M$ and $e \in E$; and
    \item The factorisation is functorial, in the sense that for
      morphisms $u,v$ with $v \after m \after e = m' \after e' \after
      u$ for $m,m' \in M$ and $e,e' \in E$, there is a 
      morphism $w$ making the following diagram commute.
      \[\xymatrix{
          \ar@{->>}^-{e}[r] \ar_-{u}[d] 
        & \ar@{-->}_-{w}[d]
        & \ar^-{v}[d] \ar@{<-<}_-{m}[l]+<2ex,0ex> \\
          \ar@{->>}_-{e'}[r]
        &  
        & \ar@{<-<}^-{m'}[l]+<2ex,0ex>
      }\]
  \end{itemize}
  A weak factorisation system is called a \emph{factorisation system}
  when the morphism $w$ above is unique.
\end{defi}
If no confusion about the (weak) factorisation system at hand can
arise, we use the notation $\twoheadrightarrow$ for morphisms in $E$, and
$\rightarrowtail$ for morphisms in $M$. Furthermore, we denote by
$M(X,Y)$ the set of morphisms in $M$ with domain $X$ and codomain
$Y$. Also, we denote by $M(X,-)$ the corresponding functor $\cat{C}
\to \Set$. 

\begin{exa}
\label{ex:factorisationsystem:posetalcategory}
  Any posetal category has a factorisation system where $E$ consists
  of all identity morphisms, and $M$ comprises all morphisms.
\end{exa}

\begin{exa}
\label{ex:factorisationsystem:VectHilb}
  An epimorphism in $\Cat{Vect}$ is a surjective linear map, a
  monomorphism in $\Cat{Vect}$ is an injective linear map. These provide
  a factorisation system for $\Cat{Vect}$.
  
  Likewise, an epimorphism in $\Cat{Hilb}$ is a continuous
  linear map with dense image, and a monomorphism in $\Cat{Hilb}$ is
  an injective continuous linear map. These provide a factorisation
  system for $\Cat{Hilb}$. 
\end{exa}
\proof
  Every epimorphism in $\Cat{Vect}$ is regular since it is the
  coequaliser of its cokernel pair~\cite[Example
  4.3.10a]{borceux}. Since the pullback of a surjective linear map is
  again a surjective linear map, the monomorphisms and (regular)
  epimorphisms form a factorisation system for
  $\Cat{Vect}$~\cite[Exercise 5.5.4]{barrwells}. The situation in
  $\Cat{Hilb}$ is analogous, except that the image first needs to be
  closed to be a genuine Hilbert space. 
\qed

The fact that the bicategory of relations is defined as a
subbicategory of the bicategory of
spans~\cite{carbonikasangianstreet,freydscedrov} inspires the
following weak factorisation system for our other running 
example, $\Cat{Rel}$. 

\begin{exa}
\label{ex:factorisationsystem:Rel}
  Call a relation $R \subseteq X \times Y$ \emph{functional}
  if $\forall_{x \in X} \exists!_{y \in Y}[ (x,y) \in R ]$, and
  \emph{oppositely functional} if $\forall_{y \in Y}\exists!_{x \in
    X}[ (x,y) \in R ]$. Denote by $M$ the collection of functional
  relations, and by $E$ the collection of oppositely functional
  relations. Then $(E,M)$ is a factorisation system for $\Cat{Rel}$.
\end{exa}
\proof
  First, isomorphisms in \Cat{Rel} are isomorphisms in \Cat{Set},
  so that these are certainly in both $E$ and $M$. Obviously, $E$ and
  $M$ are closed under composition.

  Secondly, any morphism $R \subseteq X \times Y$ of \Cat{Rel} factors
  as $R = m \after e$ for 
  \begin{align*}
    e & = \{ (x,(x,y)) \;\colon x \in X, y \in Y \mid (x,y) \in R \}
          \subseteq X \times R, \\
    m & = \{ ((x,y),y) \;\colon x \in X, y \in Y
          \mid (x,y) \in R \} \subseteq R \times Y.
  \end{align*}
  with $e \in E$ and $m \in M$.

  Thirdly, we show that the factorisation is functorial. Assume 
  \[\xymatrix{
          X \ar@{->>}^-{e}[r] \ar_-{U}[d] 
        & R
        & Y \ar^-{V}[d] \ar@{<-<}_-{m}[l]+<3ex,0ex> \\
          X' \ar@{->>}_-{e'}[r]
        & R' 
        & Y' \ar@{<-<}^-{m'}[l]+<3ex,0ex>
  }\]
  for $e,e' \in E$ and $m,m' \in M$. Then
  \[
    W = \{ ((x,y),(x',y')) \in R \times R' 
           \mid (x,x') \in U, (y,y') \in V\}
  \]
  is the unique relation between $R$ and $R'$ making both squares
  commute.   
\qed

\subsection{Compactly presentable objects}

The following observation is a combination of the notions of
compactness of objects and finite presentability that did not coincide
in $\Cat{Rel}$. Since the `monomorphisms' in
Example~\ref{ex:factorisationsystem:Rel} are functions,
$M(X,-)$ preserves \blue{bounded} directed colimits in $\Cat{Rel}$ if and only if $X$
is a finite set by
Example~\ref{ex:finitelypresentableobject:SetVectHilb}. This property
that we name `compact presentability' is now lifted to a definition,
because it turns out to be exactly what we need in
Section~\ref{sec:compactlyaccessiblecategories}.

\begin{defi}
\label{def:compactlypresentableobject}
  A compact object $X$ in a symmetric monoidal \blue{normed} category $\cat{C}$ is
  said to be \emph{compactly presentable}\footnote{The terminology is
  slightly unfortunate, because by the notation in the
  literature~\cite{adamekrosicky,gabrielulmer}  
  it might suggest that $\cat{C}(X,-)$ preserves colimits of
  $\lambda$-directed diagrams for a `compact cardinal' $\lambda$. 
  Although Definition~\ref{def:compactlypresentableobject}'s \textit{raison
  d'{\^e}tre} is to ensure the existence of the functor $(-)^*$ in
  Section~\ref{sec:compactlyaccessiblecategories}, we refrain from a
  notational name like `star-presentable object'. Likewise, `locally
  compact object' has other connotations.} with respect to a weak
  factorisation system $(E,M)$, when $M(X,-)$ preserves \blue{bounded} directed compact
  colimits.  
\end{defi}

Explicitly, a compact object $X$ is compactly presentable (with
respect to a weak factorisation system) when for any \blue{bounded} directed compact diagram
$D:\cat{J} \to \cat{C}$, any \blue{bounded} colimit cocone $d_j:D(j) \to Y$ and any morphism
$m:X \rightarrowtail Y$ in $M$, there are $j \in \cat{J}$ and a 
morphism $n:X \rightarrowtail D(j)$ in $M$ such that $m = d_j \after
n$. Moreover, this morphism $n$ is essentially unique, in the sense
that if $m=d_j \after n = d_j \after n'$, then $D(j \to j') \after n =
D(j \to j') \after n'$ for some $j' \in \cat{J}$. 
\[\xymatrix{
    D(1) \ar[r] \ar_-{d_1}[dr] 
  & D(2) \ar[r] \ar_-{d_2}[d] 
  & D(3) \ar[r] \ar_-{d_3}[dl]
  & \cdots \ar[r] 
  & D(j) \ar[r] \ar_-{d_j}[dlll] \ar@{<--<}^-{n}[d]+<0ex,3ex>
  & \cdots \\
  & Y \ar@{<-<}_-{m}[rrr]+<-3ex,0ex> & & & X 
}\]
Notice that compact presentability is a strictly stronger
notion than compactness of objects. This might be surprising because
the former depends on the structure of a weak factorisation system
whereas the latter does not. However, this is resolved by noting that
the definition of compact presentability explicitly includes the
clause that the object must be compact. 
Also, compact presentability is strictly weaker than finitely
presentable and compact, because only composition with `monomorphisms'
is required to preserve certain colimits, instead of composition with
all morphisms. This is clearly exhibited when we consider which 
objects are compactly presentable in our example categories. 

\begin{exa}
\label{ex:compactlypresentableobject:posetalmonoid}
  In a posetal category induced by an ordered commutative monoid as in
  Example~\ref{ex:compactobject:posetalgroup}, with the factorisation
  system of Example~\ref{ex:factorisationsystem:posetalcategory},
  an object $X$ is compactly
  presentable when it is invertible and in case it is smaller than a
  supremum of some directed set $D$ of invertible elements, it is
  already smaller than some element of $D$.  
\end{exa}
\proof
  Since the `monomorphisms' in the factorisation system of
  Example~\ref{ex:factorisationsystem:posetalcategory} are all
  morphisms, compactly presentable in this case coincides with compact
  and finitely presentable. The result thus follows by substituting
  Example~\ref{ex:finitelypresentableobject:posetalgroup} into
  Definition~\ref{def:compactlypresentableobject}. 
\qed

\begin{exa} 
\label{ex:compactlypresentableobject:Rel}
  In $\Cat{Rel}$, with the factorisation system of
  Example~\ref{ex:factorisationsystem:Rel}, the compactly presentable
  objects are the finite sets. 
\end{exa}
\proof
  Since the `monomorphisms' in the factorisation system in
  $\Cat{Rel}$ of Example~\ref{ex:factorisationsystem:Rel} are functions,
  an object in $\Cat{Rel}$ is compactly presentable precisely when it is
  finitely presentable in $\Cat{Set}$, which happens precisely when it
  is a finite set by
  Example~\ref{ex:finitelypresentableobject:SetVectHilb}.  
\qed

\begin{exa}
\label{ex:compactlypresentableobject:VectHilb}
  In $\Cat{Vect}$ and $\Cat{Hilb}$, with the factorisation system of
  Example~\ref{ex:factorisationsystem:VectHilb} the compactly
  presentable objects are the finite-dimensional spaces. 
\end{exa}
\proof
  Let $X$ be a finite-dimensional vector space, $D:\cat{J} \to
  \Cat{Vect}$ a directed compact diagram, and $f:X \to \colim(D)$
  an injective linear map. Since we can choose a finite basis for $X$,
  also $\mathrm{Im}(f)$ is finite-dimensional. Hence, by induction,
  $\mathrm{Im}(f)$ can be written as the span of a finite number of
  basis vectors of $\colim(D)$, for any basis of $\colim(D)$. Since
  these basis vectors must be in some $D(j)$, so is their span, and
  thus $f$ factors through $D(j)$, essentially uniquely.

  Conversely, if $f$ factors through a finite-dimensional space, $X$
  must be finite-dimensional, since $\mathrm{rank}(f)=\dim(X)$ because
  $f$ is injective.
  The situation in $\Cat{Hilb}$ is analogous.
\qed

Example~\ref{ex:compactlypresentableobject:posetalmonoid} shows that
the object $I$ is not necessarily compactly presentable: a counterexample
is the category induced by $(\field{Q},+,\leq)$, since we have $0 \leq
\sup_n(-\frac{1}{n})$, but of course $0 \not\leq -\frac{1}{n}$ for all $n$. 
Hence the compactly presentable objects do not form a monoidal
subcategory in general. However, we can ensure that the tensor product
of compactly presentable objects is again compactly presentable by the
following assumptions on the (weak) factorisation system: $\eta_X \in M$ and
$\varepsilon_X \in E$ for all $X \in \cat{C}\cpt$, and $M$ is closed
under tensor products. 

\section{Compactly presentable categories and \\ compactly accessible
  categories} 
\label{sec:compactlyaccessiblecategories}

The main idea of this article is very simple. To overcome the limitation
of finite-dimen\-sionality inherent in compact categories, we consider
categories in which every object is a \blue{bounded} directed colimit of \blue{a diagram of monomorphisms between} compact
objects. 

In $\Cat{Vect}$, this is an extension of choosing a basis for every
vector space: if $(e_n)$ is a (well-ordered) basis for $X$, then $X$
is the colimit of the totally ordered diagram 
\[\xymatrix{
  \spn(e_1) \ar[r] & \spn(e_1,e_2) \ar[r] & \spn(e_1,e_2,e_3) \ar[r] &
  \cdots  
}\]
where the morphisms are the obvious inclusions. Conversely, not every
totally ordered diagram provides a basis for its colimit vector space,
even if the constituent objects' dimension increases by one.
However, every vector space is the directed colimit of its
finite-dimensional subspaces, even if the dimension of the vector
space is uncountable. For example, consider the free complex vector
space $V=F(\field{R})$ on the basis $\field{R}$. Let $V_B = \{
\varphi: \field{R} \to \field{C} \mid \varphi(x) \neq 0 \Rightarrow x
\in B \}$ for $B \in \powerset_{\mathrm{fin}}(\field{R})$ be the
finite-dimensional subspaces. Then $V = F(\field{R}) = 
F(\colim_B B) = \colim_B F(B) = \colim_B V_B$.
Thus the slippery cardinality issue surrounding colimits of
bases is defused by the information relating different subspaces
encoded in the diagram of all finite-dimensional
subspaces. Nevertheless, the choice of a basis is a good intuition for
a directed colimit of compact objects. 

Since we will ultimately consider such categories that moreover have
an involution on the entire category (see
Section~\ref{sec:daggercompactlyaccessiblecategories}), we might as 
well consider coherence of the choice-of-duals with colimits.
For this, demanding colimits of compact objects is not enough --- it
turns out we need every object to be a directed colimit of compactly
presentable objects to construct a choice-of-duals-functor on the
entire category. 

\subsection{(Directed) \blue{bounded} colimits}

Before we can postulate every object to be a (directed) \blue{bounded} colimit of
some well-behaved kind of objects, a natural first requirement is that
the category must have all (directed) \blue{bounded} colimits. Fortunately, our
running example categories satisfy this.

In a posetal category, colimits correspond to suprema, and limits to
infima. Hence the category of
Example~\ref{ex:compactobject:posetalgroup} has directed limits and
codirected limits precisely when its partial order structure has
directed infima and directed suprema. It is complete and cocomplete
when it is a complete lattice ordered commutative monoid. 

\begin{lem} 
\label{lem:colimits:Rel}
  $\Cat{Rel}$ has \blue{bounded} directed colimits of functions and codirected limits of oppositely functional relations, but \blue{not all directed colimits and codirected limits}, and is not complete or cocomplete.
\end{lem}
\proof
  It suffices to show that $\Cat{Rel}$ has colimits of totally 
  ordered diagrams~\cite[Corollary~1.7]{adamekrosicky}. 
  Let $R_n \subseteq X_n \times X_{n+1}$ be such a chain.
  Put $X_n' = \{ x \in X_n \mid \exists_{y \in X_{n+1}} . (x,y) \in
  R_n\}$, and $X = \coprod_n X_n' \slash \mathop{\sim}$, where $\sim$ is the
  smallest equivalence relation such that $x \mathop{\sim} y$ when $(x,y) \in
  R_n$. Then $S_n = \{(x,[x]) \in X_n' \times X \mid x \in X_n'\}
  \subseteq X_n \times X$ is a cocone. To show that it is
  universal, suppose that $T_n \subseteq X_n \times Y$ is another
  cocone. Then $T_n \subseteq X_n'\times Y$, for if some $(x,y) \in
  T_n$ with $x \not\in X_n'$, then $(x,y) \not \in T_{n+1} \after R_n$
  contradicts the fact that $T_n$ forms a cocone. Define $T \subseteq X
  \times Y$ by $T = \{([x],y) \in X \times Y \mid x \in X_n' \mid
  (x,y) \in T_n\}$; this is well-defined since $T_n$ is a cocone \blue{and $R_n$ are functional}. 
  Moreover,
  \begin{align*}
        T \after S_n
    & = \{(x,y) \in X_n \times Y \mid x \in X_n' \wedge ([x],y) \in T \} \\
    & = \{(x,y) \in X_n \times Y \mid x \in X_n' \wedge (x,y) \in T_n\} \\ 
    & = T_n,
  \end{align*}
  whence $T$ is a mediating morphism. Finally, it is the unique such
  relation, since if also $T' \after S_n = T_n$, then
  $([x],y) \in T'$ for $(x,y) \in X_n' \times Y$ iff $(x,y) \in
  T_n$, so $T'=T$.
  Hence $X$ is a \blue{bounded} colimit and $S_n$ a \blue{bounded} colimiting cocone.    

  However, \blue{$\cat{Rel}$ does not have all codirected limits~\cite{milius:colimitsrel}.
  Also,} $\Cat{Rel}$ lacks equalizers. To see this, consider the 
  sets $X=\{0,1\}$ and $Y=\{0\}$, and the parallel relations 
  $R = X \times Y$ and $S= \{ (0,0) \} \subseteq X \times Y$. Their
  equaliser must be contained in $T=\{(0,0)\} \subseteq \{0\} \times
  X$.  Now $T' = \{0\} \times X$ also satisfies $R \after T' = S
  \after T'$, but does not factor through any subrelation of $T$. 

  The fact that $\Cat{Rel}$ is a self-dual category establishes the
  statements about codirected limits and completeness.
\qed

\begin{lem} 
\label{lem:colimits:Vect}
  $\Cat{Vect}$ is complete and cocomplete.
\end{lem}
\proof
  The category $\Cat{Vect}$ is algebraic, \ie
  monadic over $\Cat{Set}$. Hence it is
  complete~\cite[Theorem~3.4.1]{barrwells} and
  cocomplete~\cite[Proposition~9.3.4]{barrwells}. 
  This is also easily seen by directly constructing products,
  coproducts, kernels and cokernels~\cite[Section~V.2]{maclane2}.    
\qed

\begin{lem} 
\label{lem:colimits:Hilb}
  $\Cat{Hilb}$ has \blue{bounded} directed colimits and \blue{bounded} codirected limits, but \blue{not all directed colimits and codirected limits}.
\end{lem}
\proof
  \blue{See also~\cite{heunen:elltwo,heunen:thesis}.}
  It suffices to show that $\Cat{Hilb}$ has colimits of \blue{bounded} totally 
  ordered diagrams~\cite[Corollary 1.7]{adamekrosicky}. Denote by 
  $\Cat{Hilb}_{\leq 1}$ the category of Hilbert spaces and contractions. 
  Define a functor $F:\Cat{Hilb} \to \Cat{Hilb}_{\leq 1}$  
  by $F(H)=H$, acting on morphisms as $F(f) = \norm{f}^{-1} \cdot f$
  if $f \neq 0$ and $F(0)=0$. One easily proves that $F$ creates
  colimits of totally ordered \blue{bounded} diagrams. Since $\Cat{Hilb}_{\leq 1}$ is
  known to have directed colimits~\cite[Example 2.3.9]{adamekrosicky},
  so does $\Cat{Hilb}$. 

  To see that $\Cat{Hilb}$ does not have all colimits, consider the
  following counterexample. Define an $\field{N}$-indexed family $H_n
  = \field{C}$ of objects of $\Cat{Hilb}$, and define $f_n:H_n
  \to \field{C}$ by $f_n(z) = n \cdot z$. These are certainly bounded
  maps since $\norm{f_n}=n$. Suppose the family $(H_n)$ had a
  coproduct $H$. Then, for all $n \in \field{N}$, the norm of the
  cotuple $f:H \to \field{C}$ of $(f_n)$ must satisfy  
  \[
    n = \norm{f_n} = \norm{f \after \kappa_n} \leq \norm{f}
    \cdot \norm{\kappa_n} = \norm{f},
  \]
  where $\kappa_n$ denotes the coproduct injection, that may be
  assumed to have unit norm. This contradicts the boundedness of
  $f$, so $\Cat{Hilb}$ is not cocomplete.
  Notice that diverging behaviour as in the above counterexample is
  excluded by directed diagrams.

  The fact that $\Cat{Hilb}$ is a self-dual category establishes the
  statements about codirected limits and completeness.
\qed

\subsection{Finitely accessible categories}

For reference we now recall the kind of categories in which every
object is a (directed) colimit of finitely presentable objects. The
next subsection will imitate this construction with
compactly presentable objects instead of finitely presentable ones.

\begin{defi} 
\label{def:locallyfinitelypresentablecategory}
  A category is called \emph{locally finitely presentable} when it is
  cocomplete and has a set\footnote{In fact, we allow a set $A$ of
  finitely presentable objects, such that every object is a directed
  colimit of objects isomorphic to an object from $A$. That is, we only
  require the full subcategory of those objects to be
  \emph{essentially small}, \ie its skeleton must be small.}
  $A$ of finitely presentable objects such that every object is a
  directed colimit of objects from $A$.
\end{defi}

\begin{defi} 
\label{def:finitelyaccessiblecategory}
  A category is called \emph{finitely accessible} when it has directed
  colimits and there is a set $A$ of finitely presentable objects such
  that every object is a directed colimit of objects from $A$.
\end{defi}

Locally finitely presentable categories are precisely the free
cocompletions of small categories~\cite[Theorem 1.46]{adamekrosicky},
and finitely accessible categories are precisely the free
cocompletions of small categories with respect to directed
colimits~\cite[Theorem 2.26]{adamekrosicky}. This might suggest that
it suffices to take the free cocompletion of a compact category with
respect to directed colimits. However, then it is not clear how to
extend the choice-of-duals-functor to the resulting accessible
category (if it is possible at all). 

\subsection{Compactly accessible categories}

As mentioned above, this subsection mimicks
Definitions~\ref{def:locallyfinitelypresentablecategory}
and~\ref{def:finitelyaccessiblecategory} using compactly presentable
objects. The reason for this adaptation will become clear in the next
subsection: it ensures that the choice-of-duals-functor extends to
the entire category. Since our main example, $\Cat{Hilb}$, is not
cocomplete but has \blue{bounded} directed colimits by Lemma~\ref{lem:colimits:Hilb},
we are mostly interested in (compactly) accessible categories, but for
completeness we also consider locally (compactly) presentable categories. 

However, first we need to require the weak factorisation system to
cooperate with compactness. Compact presentability of objects already
takes into account the `monomorphisms', and the next definition fixes
coherence with the `epimorphisms'.  

\begin{defi} 
\label{def:factorisationsystem:compactlypresentable}
  A (weak) factorisation sytem is called \emph{compactly presentable}
  if quotients preserve compact presentability, that is, if $X
  \twoheadrightarrow Y$ and $X$ is compactly presentable, then so is
  $Y$. 
\end{defi}

The (weak) factorisation systems for posetal categories, $\Cat{Rel}$,
$\Cat{Vect}$ and $\Cat{Hilb}$ we met in 
Examples~\ref{ex:factorisationsystem:VectHilb} 
and~\ref{ex:factorisationsystem:Rel} are all easily seen to be
compactly presentable. If $X$ is a finite set and there is a
surjection onto $Y$, then surely $Y$ is finite. Likewise, if $X$ is a
finite-dimensional vector space and there is a surjection onto $Y$,
then also $Y$ must be finite-dimensional.

\begin{defi}  
\label{def:category:locallycompactlypresentable}
  A \blue{normed} category $\cat{C}$ is called \emph{locally compactly presentable} if
  \begin{itemize}
    \item it is symmetric monoidal;
    \item it has \blue{bounded} compact limits and \blue{bounded} compact colimits;
    \item it is equipped with a compactly presentable weak
      factorisation system; and 
    \item it has a set $A$ of compactly presentable objects such that
      every object is a \blue{bounded} directed colimit of objects of $A$.
  \end{itemize}
\end{defi}

\noindent Since every set is a directed colimit (in $\Cat{Rel}$) of the diagram
of its finite subsets (ordered by
inclusion)\footnote{Notice that, up-to-isomorphism, there is only a set
of finite sets (namely $\field{N}$).}, we see that $\Cat{Rel}$,
equipped with the factorisation system of
Example~\ref{ex:compactlypresentableobject:Rel}, is a compactly
presentable category by collecting earlier results. \newpage

\begin{defi}  
\label{def:category:compactlyaccessible}
  A \blue{normed} category $\cat{C}$ is called \emph{compactly accessible} if
  \begin{itemize}
    \item it is symmetric monoidal;
    \item it has \blue{bounded} directed compact colimits and \blue{bounded} codirected compact limits;
    \item it is equipped with a compactly presentable weak
      factorisation system; and 
    \item it has a set $A$ of compactly presentable objects such that
      every object is a \blue{bounded} directed colimit of objects of $A$.
  \end{itemize}
\end{defi}

\noindent Of course, every compactly presentable category --- \eg $\Cat{Rel}$
--- is a compactly accessible category. But also $\Cat{Vect}$ and
$\Cat{Hilb}$ (and their 
generalisations indicated in Examples~\ref{ex:compactobject:Vect} and
~\ref{ex:compactobject:Hilb}), equipped with their canonical
factorisation systems (see
Example~\ref{ex:factorisationsystem:VectHilb}) \blue{and norms}, are examples of
compactly accessible categories. Hence the previous definition at
least succeeds in overcoming the limitation of finite-dimensionality. 

We also see that the posetal category induced by an directed-complete
ordered Abelian group of Example~\ref{ex:compactobject:posetalgroup}
is a compactly accessible category precisely if every element is a
directed supremum of the compact elements below it, \ie when
the Abelian group is in fact ordered by an algebraic
domain~\cite{abramskyjung}.  

\subsection{Properties of compactly accessible categories}

Compactly accessible categories inherit some of the pleasant
properties from compact categories, and others only partly. The
choice-of-duals-functor, that is arguably the most important feature
of a compact category, extends canonically. We first define the
construction on morphisms, and then prove it to be functorial. 

\begin{defi}
\label{def:dualmorphism}
  Let $f:X\to Y$ be a morphism in a compactly accessible
  category $\cat{C}$. Pick directed compactly presentable diagrams
  $C:\cat{I}\to\cat{C}$ and  
  $D:\cat{J}\to\cat{C}$ with \blue{bounded} colimit cocones $c_i:C(i) \to X$ and
  $d_j:D(j) \to Y$. Let \blue{bounded} limit cones $c_i^*:X^* \to C^*(i)$ and
  $d_j^*:Y^* \to D^*(j)$ be given. We define a morphism $f^*:X^* \to
  Y^*$ as follows. 

  Every $f \after c_i$ factors as $\smash{\xymatrix@1{C(i) \ar@{->>}^-{f_i}[r] & X_i
  & Y \ar@{<-<}_-{m_i}[l]+<3ex,0ex>}}$. Because $C(i)$ is compactly
  presentable, so is $X_i$, and hence there is a $j_i \in \cat{J}$
  such that $m_i$ 
  factors as $\smash{\xymatrix@1{X_i \ar^-{n_i}[r] & D(j_i)
  \ar^-{d_{j_i}}[r] & Y}}$. Because of the functorial
  property of the weak factorisation system and directedness of $D$, the
  morphisms $f_i^* \after n_i^* \after d_{j_i}^*$ form a \blue{bounded} cone from
  vertex $Y^*$ to $C^*$. Define $f^*$ to be the unique mediating
  morphism $Y^* \to X^*$.   
  \[\xymatrix{
    C(i) \ar^-{c_i}_-{\colim}[dd] \ar@{->>}^-{f_i}[dr] 
    & & D(j_i) \ar^-{\colim}_-{d_{j_i}}[dd] 
    & C^*(i) & & D^*(j_i) \ar_-{n_i^*}[dl] \\
    & X_i \ar@{-->}^-{n_i}[ur] & & & X_i^* \ar_-{f_i^*}[ul] \\
    X \ar_-{f}[rr] & & Y \ar@{<-<}_-{m_i}[ul]+<2ex,-2ex> 
    & X^* \ar_-{c_i^*}^-{\lim}[uu] 
    & & Y^* \ar_-{\lim}^-{d_{j_i}^*}[uu] \ar@{-->}^-{f^*}[ll]
  }\]
\end{defi}

\begin{lem}
\label{lem:dualmorphismfunctorial}
  Let $\xymatrix@1{X \ar^-{f}[r] & Y \ar^-{g}[r] & Z}$ be morphisms in
  a compactly accessible category $\cat{C}$. Pick directed compactly
  presentable diagrams 
  $C:\cat{I}\to\cat{C}$, $D:\cat{J}\to\cat{C}$ and
  $E:\cat{K}\to\cat{C}$ with \blue{bounded} colimit cocones $c_i:C(i) \to X$, 
  $d_j:D(j) \to Y$, and $e_k:E(k) \to Z$. Let \blue{bounded} limit cones $c_i^*:X^* \to
  C^*(i)$, $d_j^*:Y^* \to D^*(j)$ and $e_k^*:Z^* \to E^*(k)$ be 
  given. Then $(g \after f)^*$, as defined above, equals $f^* \after g^*$.  
\end{lem}
\proof
  According to the construction in the previous definition, we get the
  following commuting diagrams.
  \[\xymatrix@C-1.5ex{
    C(i) \ar_-{c_i}[dd] \ar@{->>}^-{f_i}[dr]
    & & D(j_i) \ar^-{d_{j_i}}[dd] \ar@{->>}^-{g_i}[dr]
    & & E(k_i') \ar^-{e_{k_i'}}[dd]
    & & C(i) \ar_-{c_i}[dd] \ar@{->>}^-{(g \after f)_i}[dr]
    & & E(k_i'') \ar^-{e_{k_i''}}[dd] \\
    & X_i \ar@{-->}_-{n_i}[ur]
    & & Y_i \ar@{-->}_-{n_i'}[ur]
    & & & & Z_i \ar@{-->}_-{n_i''}[ur] \\
    X \ar_-{f}[rr] 
    && Y \ar_-{g}[rr] \ar@{<-<}_-{m_i}[ul]+<2ex,-2ex>
    && Z \ar@{<-<}_-{m_i'}[ul]+<2ex,-2ex>
    && X \ar_-{g \after f}[rr] 
    && Z \ar@{<-<}_-{m_i''}[ul]+<2ex,-2ex>
  }\]
  The functoriality of the weak factorisation system provides a morphism
  $w_i$ making the following diagram commute.
  \[\xymatrix@C+2ex{
    C(i) \ar_-{c_i}[dd] 
               \ar@{->>}^-{f_i}[r] \ar@{->>}_-{(g \after f)_i}[drr]
    & X_i \ar^-{n_i}[r] & D(j_i) \ar@{->>}^-{g_i}[dr] \\
    && Z_i \ar@{-->}_-{w_i}[r] & Y_i \\
    X \ar_-{f}[rr] && Y \ar_-{g}[rr]
    && Z \ar@{<-<}_-{m_i'}[ul]+<2ex,-2ex>
         \ar@{<-<}^-{m_i''}[ull]+<2ex,-2ex>
  }\]
  Because $E$ is a \blue{bounded} directed diagram, there exist $k_i \in \cat{K}$ and
  morphisms $a':k_i' \to k_i$ and $a'':k_i'' \to k_i$ of
  $\cat{K}$. So the following diagram commutes. 
  \[\xymatrix@R-2ex@C+2ex{
    Z_i \ar^-{n_i''}[r] \ar_-{w_i}[dd] 
    & E(k_i'') \ar^-{E(a'')}[r] & E(k_i) \ar^-{e_{k_i}}[dr] \\
    & & & Z \\
    Y_i \ar_-{n_i'}[r] & E(k_i') \ar_-{E(a')}[r] & E(k_i) \ar_-{e_{k_i}}[ur]
  }\]
  Hence we get compatible cones $Z^* \to C^*$, as in the following diagram.
  \[\xymatrix@C+2ex{
    C^*(i) & Z_i^* \ar_-{(g \after f)_i^*}[l] & E^*(k_i'')
    \ar_-{n_i''^*}[l] & E^*(k_i) \ar_-{E(a'')^*}[l] \\ 
    & & & & Z^* \ar_-{e_{k_i}^*}[ul] \ar^-{e_{k_i}^*}[dl] \\
    X_i^* \ar^-{f_i^*}[uu] & Y_i^* \ar^-{n_i^* \after g_i^*}[l]
    \ar_-{w_i^*}[uu] & E^*(k_i') \ar^-{n_i'^*}[l] & E^*(k_i)
    \ar^-{E(a')^*}[l] 
  }\]
  Thus by uniqueness of the mediating morphism, $(g \after f)^* =
  f^* \after g^*$. 
\qed

\begin{thm}
\label{thm:starfunctor}
  There is a canonical functor $(-)^*:\cat{C}\op\to\cat{C}$ on any
  compactly accessible category $\cat{C}$, extending that on $\cat{C}\cpt$.
\end{thm}
\proof
  An easy diagram chase shows directly that the construction in Definition
  \ref{def:dualmorphism} satisfies $\idmap^*=\idmap$. Combining this
  with Lemma~\ref{lem:dualmorphismfunctorial}, we see that $\colim(C)
  \cong \colim(D)$ for compactly presentable \blue{bounded} directed diagrams $C$ and
  $D$ implies that $\lim(C^*) \cong \lim(D^*)$. Hence picking one
  representative $X^*$ of each isomorphism 
  class $\lim(D^*)$ where $X\cong\colim(D)$ provides an action
  $(-)^*:\cat{C}\op\to\cat{C}$ on objects. Definition 
  \ref{def:dualmorphism} subsequently gives an action on morphisms,
  and Lemma \ref{lem:dualmorphismfunctorial} shows that this is indeed
  functorial. 

  If we ensure that the choice of representatives $X^*,Y^*$ coincides
  with the choice of duals for $X,Y \in \cat{C}\cpt$, then the
  situation for a morphism 
  $f:X \to Y$ collapses, so that the $f^*$ of Definition
  \ref{def:dualmorphism} indeed coincides with the $f^*$ of
  Proposition \ref{prop:dualmorphismcpttocpt}. After all, 
  $C,D:\cat{1} \to \cat{C}$ with $C(*)=X$ and $D(*)=Y$ are compact
  directed \blue{bounded} diagrams. 
\qed

In $\Cat{Vect}$, equipped with its usual factorisation system (of
Example~\ref{ex:factorisationsystem:VectHilb}), the functor $(-)^*$ of
the previous theorem maps an object to its usual dual vector space
(and a morphism to its usual dual). Hence, the dual-space functor of
vector spaces is entirely determined when a choice of dual spaces of
just the finite-dimensional vector spaces (and a factorisation system)
has been fixed.  

However, by allowing infinite-dimensionality in compactly accessible
categories, we also partly lost some properties of compact
categories. For example, the functor $(-)^*$ is no longer an equivalence.

\begin{prop}
  The isomorphism $X \cong X^{**}$ holds for compact objects $X$ in a
  compactly accessible category $\cat{C}$, but not for any object.
\end{prop}
\proof
  For compact objects we have
  Proposition~\ref{prop:compactcategory:starisequivalence}.
  As a counterexample for non-compact objects, we already saw 
  in Example~\ref{ex:compactobject:Vect} that an
  infinite-dimensional vector space is not isomorphic to its double
  dual by a cardinality argument. 
\qed

Unfortunately this entails that the choice-of-duals-functor is no longer 
necessarily involutive up to isomorphism outside the compact part of the
category. However, for the present purpose this is not a major issue,
because the `essence' of a quantum protocol resides in the compact
part; the cocompletion aspect is only used because the
dimension is not a priori bounded. 

However, the canonical factorisation system in
$\Cat{Hilb}$ (see Example~\ref{ex:factorisationsystem:VectHilb})
provides a canonical extension of the choice-of-duals functor that
\emph{is} an equivalence. It is remarkable that such a functor can
be derived from the axiomatic structure of compactly accessible categories.

Likewise, a compactly accessible
category is no longer a tensored *-category~\cite{mueger} in the sense
that the tensor does not necessarily cooperate with the
choice-of-duals-functor outside the compact part of the category.

\begin{prop} 
  If $X$ or $Y$ is compact, then $(X \tensor Y)^* \cong X^* \tensor
  Y^*$, but this isomorphism does not hold in general in a compactly
  accessible category. 
\end{prop}
\proof
  Without loss of generality, assume $X$ to be compact. Let a
  directed compactly presentable diagram $D:\cat{J}\to\cat{C}$ with
  colimit $Y$ be given. Since $X^*$ is also compact, we have by
  Proposition~\ref{prop:propertiesofcompactobjects}(e) that
  \begin{align*}
        (X \tensor Y)^*
    & = (X \tensor \colim_j(D(j)))^* \\
    & = (\colim_j(X \tensor D(j)))^* \\
    & = \lim_j (X^* \tensor D^*(j)) \\
    & = X^* \tensor (\lim_j(D^*(j))) \\
    & = X^* \tensor Y^*. 
  \end{align*}

\noindent However, for infinite-dimensional vector spaces $X$ and $Y$ it is
  not necessarily true that $(X \tensor Y)^* \cong X^* \tensor
  Y^*$. 
\qed

Again, it is interesting to remark that the previous proposition does
hold for all Hilbert spaces $X$ and $Y$, since every Hilbert space is
naturally isomorphic to its dual (by the Riesz representation theorem).

\section{Dagger compactly accessible categories}
\label{sec:daggercompactlyaccessiblecategories}

Although the choice-of-duals is arguably the most important
feature, to model quantum protocols one needs a compact category to have a
second involutive structure coherent with
choice-of-duals~\cite{abramskycoecke,selinger}. In the prime example 
category of finite-dimensional Hilbert spaces, this second structure
provides complex conjugation, whereas the choice-of-duals accounts
for transposition of matrices. 

\begin{defi}
  A \emph{dagger category} is a category $\cat{C}$ equipped with an
  involutive, identity-on-objects functor $(-)^\dag : \cat{C}\op \to
  \cat{C}$.
  \blue{A \emph{normed dagger category} additionally satisfies $\|f^\dag\|=\|f\|$.}
\end{defi}

All kinds of terminology transfers from $\Cat{Hilb}$ to any dagger
category. For example, a morphism $f$ in a dagger category is called 
an \emph{isometry} when $f^\dag \after f=\idmap$, and \emph{unitary}
when furthermore $f \after f^\dag=\idmap$.

\begin{defi} 
  A \emph{dagger symmetric monoidal category} is a symmetric monoidal
  category $\cat{C}$ that is simultaneously a dagger category, such that 
  \[
    (f \tensor g)^\dag = f^\dag \tensor g^\dag,
  \]
  and the associativity, left-unit, right-unit and symmetry monoidal
  structure isomorphisms are unitary.
\end{defi}

Examples of dagger symmetric monoidal categories are $\Cat{Rel}$ and
$\Cat{Hilb}$. In $\Cat{Rel}$, the dagger structure is given on
morphisms by $R^\dag = \{(y,x) \;\colon (x,y) \in R\}$. The dagger
structure in $\Cat{Hilb}$ is provided by the Riesz representation
theorem: for a morphism $f:X \to Y$ of $\Cat{Hilb}$, there is a unique
morphism $f^\dag : Y \to X$ satisfying $\inprod{f(x)}{y} =
\inprod{x}{f^\dag(y)}$ for all $x \in X$ and $y \in Y$~\cite[Theorem
2.4.2]{kadisonringrose}. 

We now adapt the definition of `dagger compact category' slightly to
encompass compactly accessible categories. First, the dagger structure
should cooperate with the weak factorisation system.

\begin{defi}
  A (weak) factorisation system $(E,M)$ in a dagger category is called
  a \emph{dagger (weak) factorisation system} when $e^\dag$ is in $M$ for
  each $e$ in $E$, and $m^\dag$ is in $E$ for each $m$ in
  $M$.\footnote{\label{footnote:factorisationsystemassquareroot}`Factorisation'
  can be taken more literally by viewing $M$ and $E$ as subcategories of
  $\cat{C}$ and saying $\cat{C} = M \after E$. A
  dagger factorisation system then resembles a square root, as $\cat{C}
  = E^\dag \after E$, or ``$E = \sqrt{\cat{C}}$''.}
\end{defi}

The (weak) factorisation systems of $\Cat{Rel}$ and $\Cat{Hilb}$ of
Examples~\ref{ex:factorisationsystem:VectHilb}
and~\ref{ex:factorisationsystem:Rel} are obviously dagger (weak)
factorisation systems with respect to the above dagger structure.

The next definition essentially states that a category
$\cat{C}$ is dagger compactly accessible when it is a dagger category
that is also compactly accessible, and $\cat{C}\cpt$ is dagger
compact~\cite{selinger}.

\begin{defi} 
  A \emph{dagger compactly accessible category} is a compactly
  accessible category $\cat{C}$ that is also a dagger category, such
  that the weak factorisation system is a dagger weak factorisation
  system, and $\sigma \after \varepsilon_X^\dag = \eta_X : I \to X^*
  \tensor X$ for all $X \in \cat{C}\cpt$, where $\sigma:X \tensor X^*
  \to X^* \tensor X$ denotes the symmetry isomorphism.
\end{defi}

Of course, our running example categories $\Cat{Rel}$ and $\Cat{Hilb}$
are both dagger compactly accessible categories, since $\Cat{Rel}\cpt$
and $\Cat{Hilb}\cpt$ are dagger compact categories.

The most important property of a dagger structure in relation to a
compact category is that the dagger functor commutes with the
choice-of-duals-functor. This provides `complex conjugation'. This
pleasant property is retained in full in dagger compactly accessible
categories. 

\begin{thm}
\label{thm:daggercommuteswithstar}
  For every morphism $f:X \to Y$ in a dagger compactly accessible
  category $f^{\dag *}=f^{*\dag}:X^* \to Y^*$ holds. 
\end{thm}
\proof
  Since $(-)^{\dag}$ is a strict involution, if $c_i:X \to C(i)$ is a
  \blue{bounded} limit cone, then $c_i^\dag:C^\dag(I) \to X$ is a \blue{bounded} colimit cocone, and
  vice versa. Moreover, since the weak factorisation system respects
  the dagger, if $f:X \to Z$ factors as $\xymatrix@1{X
  \ar@{->>}^-{e}[r] & Y & Z \ar@{<-<}_-{m}[l]+<2ex,0ex>}$, then
  $f^\dag:Z \to X$ factors as $\xymatrix@1{Z \ar@{->>}^-{m^{\dag}}[r]
  & Y & X \ar@{<-<}_-{e^{\dag}}[l]+<2ex,0ex>}$. Hence in the defining
  diagrams of $f^{*\dag}$ and $f^{\dag*}$ below,
  \[\xymatrix{
    C^*(i) \ar_-{c_i^{*\dag}}[dd] \ar^-{f_i^{*\dag}}[dr] 
    & & D^*(j_i) \ar^-{d_{j_i}^{*\dag}}[dd]  & 
    C^*(i_j) \ar^-{(n'_i)^*}[dr] & & D^*(j) \\
    & X^*_i \ar^-{n_{j_i}^{*\dag}}[ur] & & & 
    X_{i_j}^* \ar^-{(f^{\dag})_j^*}[ur] \\
    X^* \ar_-{f^{*\dag}}[rr] & & Y^* & 
    X^* \ar^-{c_{i_j}^*}[uu] \ar_-{f^{\dag *}}[rr] & & Y^* \ar_-{d_j^*}[uu]
  }\]
  one has that $d_{j_i}^{*\dag} \after n_{j_i}^{*\dag} \after
  f_i^{*\dag}$ and $d_j^{*\dag} \after (f^\dag)_j^* \after (n'_i)^*$
  form the same \blue{(bounded)} cocone. Since also $c_i^{*\dag}$ and $c_{i_j}^{*\dag}$
  form the same \blue{(bounded)} cocone, the mediating morphisms $f^{*\dag}$ and
  $f^{\dag *}$ coincide.
\qed

By the previous theorem, there is a covariant functor
$(-)_*:\cat{C} \to \cat{C}$ determined by $X_*=X^*$
on objects and acting as $f_*=f^{*\dag}=f^{\dag*}$ on
morphisms~\cite[Definition~2.9]{selinger}. In $\Cat{Hilb}$ with its
usual factorisation system (of
Example~\ref{ex:factorisationsystem:VectHilb}), it maps a morphism to
its complex conjugate.

\subsection{Structure or property?}
\label{subsec:structureorproperty}

As a technical intermezzo, let us consider the status of compact
accessibility: is it a structure or a property? 
A compactly accessible category requires a compactly presentable weak
factorisation system. Initially it is not clear that this will
uniquely exist. This makes compact accessibility into a structure,
whereas compactness is a property.

First, notice that a compactly presentable factorisation system always
exists. Any symmetric monoidal category has a compactly presentable
factorisation system in which $E$ is comprised of all isomorphisms and
$M$ consists of all morphisms. 

This immediately shows that a compactly presentable factorisation
system is not unique. A forteriori, this shows that the notion of
compact presentability of objects is not independent of the chosen
factorisation system: in $\Cat{Rel}$ with the above factorisation
system, every object is compactly presentable, whereas in $\Cat{Rel}$
with the factorisation system of
Example~\ref{ex:factorisationsystem:Rel}, only the finite sets are. 

However, although these intermediate considerations might not be
independent of the factorisation system used, the canonical extension
of the choice-of-duals functor is, as soon as it is on objects. The
following proposition states this rigorously.

\begin{prop}
  Let $(E,M)$ and $(E',M')$ both be compactly presentable
  weak factorisation systems for a symmetric monoidal category
  $\cat{C}$. 
  If objects $X,Y$ are compactly presentable with respect to
  both factorisation systems, then $X^*,Y^*$ are independent of the
  factorisation system used. 
  Moreover, in that case $f^*$ is independent of the factorisation
  system used for any morphism $f:X \to Y$.
\end{prop}
\proof
  The claim on objects is just a reformulation of the hypothesis.
  Let us consider the claim on morphisms in the notation of
  Definition~\ref{def:dualmorphism}: suppose $c_i:C(i) \to X$ and $d_j
  : D(j) \to Y$ are colimit cones, and that $f \after c_i$ factors as
  $m_i \after e_i$ and $m_i' \after e_i'$ in both factorisation
  systems, respectively. Then there are $n_i$ and $n_i'$ such that
  $m_i = d_{j_i} \after n_i$ and $m_i' = d_{j_i'} \after n_i'$.
  Since $D$ is directed, there is a $d_i$ with $d_i \after n_i = m_i$
  and $d_i \after n_i' = m_i'$. So $d_i \after n_i \after e_i = f
  \after c_i = d_i \after n_i' \after e_i'$, whence $e_i^* \after
  n_i^* \after d_i^*$ and $e_i'^* \after n_i'^* \after d_i^*$ form
  compatible cones, and both mediating morphisms $f^*$ coincide.
\qed

With the intuition of
footnote~\ref{footnote:factorisationsystemassquareroot}, 
one might suspect that a factorisation system is unique in the
presence of a dagger functor.
In any factorisation system, each of the classes $E$ and $M$
is determined by the other via so-called orthogonality 
$E=M^\perp$~\cite[Proposition~5.5.3]{borceux}. 
Thus, the larger $E$ is, the smaller $M$ can be.
But compatibility with the dagger functor moreover requires $E=M^\dag$,
guaranteeing that $E$ and $M$ balance each other in size.
However, here is an example of a dagger category with two different
dagger compactly presentable factorisation systems. 
For any category $\cat{C}$, the cofree dagger category
$\cat{C}_\leftrightarrows$ has the same objects; a morphism $X \to Y$
in $\cat{C}_\leftrightarrows$ consists of a pair of morphisms
$f_\leftarrow : Y \to X$ and $f_\rightarrow : X \to Y$ of
$\cat{C}$, with $(f_\leftarrow,f_\rightarrow)^\dag =
(f_\rightarrow,f_\leftarrow)$. All kinds of structures lift through
this construction. If $\cat{C}$ is symmetric monoidal, so is
$\cat{C}_\leftrightarrows$. An object in $\cat{C}$ is compact iff it
is in $\cat{C}_\leftrightarrows$. An object in $\cat{C}$ is compactly
presentable iff it is in $\cat{C}_\leftrightarrows$. A factorisation
system for $\cat{C}$ lifts to a dagger factorisation system for
$\cat{C}_\leftrightarrows$. Thus, a compactly presentable
factorisation system for $\cat{C}$ lifts to a dagger compactly
presentable factorisation system for $\cat{C}_\leftrightarrows$. 
Hence the above example in $\Cat{Rel}$ provides an example of a dagger
category $\Cat{Rel}_\leftrightarrows$ with two different dagger
compactly presentable factorisation systems.

Still, in all our example categories the dagger compactly presentable
factorisation systems had a very canonical feel to them.
All in all, dagger factorisation systems suggest themselves as a
worthy subject of further study in their own right.

\subsection{Classical structures and measurements}
\label{subsec:classicalstructuresandmeasurements}

Finally, we need to model measurements in our semantics. These
can be dealt with categorically~\cite{coeckepavlovic} --- the
following definitions recall the necessary notions, adapted to the
compactly accessible setting.

\begin{defi}
\label{def:classicalstructure}
  An object $C$ in a dagger compactly accessible category
  $\cat{C}$ is said to be a \emph{classical structure} when it is
  equipped with a commutative comonoid structure  
  \[\xymatrix@1{
    C \tensor C & C \ar_-{\delta}[l] \ar^-{\epsilon}[r] & I
  }\]
  in which $\delta$ is an isometry, that moreover
  satisfies $\delta \after \delta^\dag = (\delta^\dag \tensor \idmap)
  \after (\idmap \tensor \delta)$.
\end{defi}

The precise meaning of the technical condition is not important
here. The idea is that $\delta$ provides a `copying' operation, and
$\epsilon$ a `deleting' operation. The definition thus
counterfactually exploits the fact that quantum data cannot be cloned
or forgotten. We remark that a classical structure
$(C,\delta,\epsilon)$ automatically satisfies
Diagrams~\eqref{eq:compactobject} with $C^*=C$, $\eta=\delta 
\after \epsilon^\dag$ and $\varepsilon=\epsilon \after \delta^\dag$. 
For more information we refer to~\cite{coeckepavlovic}.

We tentatively call an object in a dagger compactly accessible
category that is not equipped with a fixed classical structure a 
\emph{quantum object}. Any infinite-dimensional Hilbert space is a
quantum object in $\Cat{Hilb}$, since it cannot carry any classical
structure as that would entail
finite-dimensionality~\cite{kock:frobenius}. We refer 
to~\cite{coeckeduncan:interactingquantumobservables} for a way
to select quantum objects representing qubits.

The type of a (demolition) measurement now is $X \to C$, for a
classical structure $C$. As in the traditional Hilbert space
formalism, we first define a basis, or projector-valued spectrum, in
which to measure. 

\begin{defi}
\label{def:measurement}
  A \emph{demolition projector-valued spectrum} on an object $X$ in a
  dagger compactly accessible category $\cat{C}$ is a morphism $p:X
  \to C$, whose codomain $C$ is a classical structure, that satisfies $p
  \after p^\dag = \idmap[C]$.
\end{defi}

In other words, a demolition projector-valued spectrum is the
adjoint of an isometry, and hence the splitting of an
idempotent~\cite{selinger2}. 

Now, a demolition measurement is nothing but a shell around a
projector-valued spectrum that `eliminates global
phases'~\cite{coeckepavlovic}. We ignore this and use measurement and
projector-valued spectrum as synonyms.

\section{Quantum key distribution, categorically}
\label{sec:quantumkeydistributioncategorically}

With dagger compactly accessible categories in place as a semantics,
the stage is now set to model the quantum key distribution protocol in
Figure~\ref{fig:e91}. 
As mentioned before, as of yet we can only model the
qualitative steps $\one$, $\two$, $\three$, $\four$ and $\seven$
categorically. In fact, the entire purpose 
of a categorical semantics is to abstract away from the quantative
details in steps $\five$ and $\six$. Though a categorical version of
inequalities like Bell's would not be superfluous, it is outside of
the scope of this article. 

\subsection{The quantum channel}

The feature of the protocol in Figure~\ref{fig:e91} that cannot be
accomodated in a dagger compact category is the possibly unbounded
need for fresh qubit-pairs. Hence, as a preparation we set up an
object from which to draw an a priori unknown number of qubit-pairs. 
Let $\cat{C}$ be a dagger compactly accessible category. Select a
quantum object $X$ in $\cat{C}$, such that $X^* \tensor X$ is
compactly presentable \blue{and $\eta \colon I \to X^* \tensor X$ is in $M$}, to represent the qubit. Define a diagram
$D:\field{N} \to \cat{C}$ by  
\begin{align*}
  D(n) & = (X^* \tensor X)^{\tensor n} = \underbrace{(X^* \tensor X) \tensor
  \cdots \tensor (X^* \tensor X)}_{n~\mathrm{times}}, \\
  D(n \to n+1) & :\; \xymatrix@1{D(n) \ar_-{\cong}[r] & D(n) \tensor I
    \ar_-{\idmap \tensor \eta}[r] & D(n+1)}
\end{align*}
This is a directed compactly presentable diagram\blue{, that is bounded by $\|\eta\|$,} and hence it has a
colimit $Z=\colim(D)$. This object $Z$ will function as a store of
qubit-pairs that are guaranteed to be fresh; it models the quantum
channel (and the index $\field{N}$ of the colimit represents
`time'). Notice that this is not possible with $X$ alone since that
object is compact.\footnote{There is a resemblance to 
type theory here: as $X$ is a `finite type', we need to have countably
many copies of it to be able to draw countably many distinct variable
letters.}  

One can now draw a fresh qubit-pair from the quantum channel $Z$ as
follows. Let $d_n:D(n) \to Z$ be a colimit cone. Then $\idmap \tensor
d_{n-1} : D(n) \to (X^* \tensor X) \tensor Z$ forms another cone to
$Z$, and hence there is a unique mediating morphism $d:Z \to (X^*
\tensor X) \tensor Z$. 

The following reasoning shows that $d:Z \to (X^* \tensor X) \tensor Z$
is in fact an isomorphism. Since $X^* \tensor X$ is a compact object, 
$(X^* \tensor X) \tensor (-)$ is cocontinuous by
Proposition~\ref{prop:propertiesofcompactobjects}(e). Hence we have:
\begin{align*}
          (X^* \tensor X) \tensor Z 
  & =     (X^* \tensor X) \tensor \colim_n((X^*\tensor X)^{\tensor n}) \\
  & \cong \colim_n((X^* \tensor X)^{\tensor(n+1)}) \\
  & \cong \colim_n((X^* \tensor X)^{\tensor n}) 
    =     Z.
\end{align*}
Moreover, the diagram $\xymatrix@1{I \ar^-{d_0}[r] & Z & (X^* \tensor X)
\tensor Z \ar_-{d^{-1}}[l]}$ is initial in the sense that for any
given diagram $\xymatrix@1{I \ar^-{f}[r] & A & (X^* \tensor X) \tensor 
A \ar_-{g}[l]}$ there is a unique mediating morphism $Z \to A$.
It is constructed via the \blue{bounded} colimit.
We can understand $Z$ as a list object with elements from
$X^* \tensor X$: these objects are usually defined as initial
algebras of the functor $1+(X^* \tensor X) \tensor (-)$, but since our
situation does not necessarily provide a coproduct we used 
cospans instead.\footnote{Naming the carrier of the initial diagram
$(X^* \tensor X)^\star$ instead of $Z$ would be apt but confusing.}

Thus $Z$ models the quantum channel, and $d:Z \to (X^* \tensor X)
\tensor Z$ represents drawing one fresh qubit-pair prepared in
Bell state.

\subsection{The categorical model of the protocol}

Having dealt with the qubit and the quantum channel, step $\one$ now
provides us with demolition measurements $m_i:X \to C$, where the
classical structure $C$ represents the bit. The protocol 
in Figure~\ref{fig:e91} can now be modelled as follows.  
\[\xymatrix{
  I \ar^-{d_0}_-{\one}[r] 
  & Z \ar^-{d^{\tensor 3n}}_-{\two,\three}[rr] 
  && (X^* \tensor X)^{\tensor 3n} \tensor Z
  \ar_-{\four}^(.33){\qquad(\bigotimes_{i=1}^{3n} m_{a_i} \tensor
    (m_{b_i})_*) \tensor \idmap}[dl] \\
  & & C^{\tensor 3n} \tensor C^{\tensor 3n} \tensor Z \ar_-{\seven}[r] \ar^-{\five,\six}[ul]
  & C^{\tensor 2n} \tensor C^{\tensor 2n} \tensor Z
}\]
Since we chose to keep classical communication
external, steps $\two$, $\five$, $\six$ and $\seven$ depend on external
events. Steps $\five$, $\six$ and $\seven$ are modeled by forgetting
the relevant information using $\epsilon$, and step $\two$ is fully
external. Thus, the protocol is 
represented by a morphism $I \to C^{\tensor 2n} \tensor C^{\tensor 2n}
\tensor Z$ with probability one: starting from nothing, Alice and Bob
each end up having $2n$ bits, and there is still the possibility of obtaining
fresh qubit-pairs on their shared quantum channel. The probabilistic
branching, and in particular the (improbable) possibility of
non-termination, could be dealt with more precisely using coalgebraic
techniques~\cite{hasuojacobssokolova,bucalorosolini}. However, the
above suffices as an illustration of the need for dagger compactly
accessible categories.  

We are now in a position to prove the correctness of the protocol
categorically, \ie to prove that Alice and Bob in
fact end up with equal key bits, without assuming anything about
the demolition measurements $m_i$ or the external choices of $a_i$ and
$b_i$. It suffices to prove this for each individual key bit that
arises from Alice and Bob using the same measurement, 
because step $\seven$ discards the other bits. Hence the correctness
of the protocol comes down to the following theorem.

\begin{thm}
  The following diagram commutes for any demolition projector-valued
  spectrum $m:X \to C$.
  \[\xymatrix@C-1.5ex{
    Z \ar@{=}[d] \ar^-{d}[r] & X^* \tensor X \tensor Z
    \ar^-{m_* \tensor m \tensor \idmap}[rr] && C^* \tensor C
    \tensor Z \ar^-{\epsilon \tensor \idmap 
    \tensor \idmap}[rr] && I \tensor C 
    \tensor Z \ar^-{\cong}[r] & C \tensor Z \ar@{=}[d] \\
    Z \ar_-{d}[r] & X^* \tensor X \tensor Z \ar_-{m_* \tensor
      m \tensor \idmap}[rr] &&
    C^* \tensor C \tensor Z \ar_-{\idmap \tensor \epsilon \tensor
      \idmap}[rr] && C \tensor I \tensor Z \ar_-{\cong}[r] & C \tensor Z
  }\]
\end{thm}
\proof
  Because $Z$ is only acted upon by the identity morphism,
  it suffices to prove commutativity of the following diagram. 
  \[\xymatrix{
    I \ar^-{\eta_X}[r] \ar@{=}[d]
  & X^* \tensor X \ar^-{m_* \tensor m}[r]
  & C^* \tensor C \ar^-{\epsilon \tensor \idmap{}}[r]
  & I \tensor C \ar^-{\cong}[r]
  & C \ar@{=}[d] \\
    I \ar_-{\eta_X}[r]
  & X^* \tensor X \ar_-{m_* \tensor m}[r]
  & C^* \tensor C \ar_-{\idmap{} \tensor \epsilon}[r]
  & C^* \tensor I \ar_-{\cong}[r]
  & C
  }\]
  First, notice that
  \[
    (m_* \tensor m) \after \eta_X
   = (m_* \tensor m) \after \name{\idmap[X]} \\
   \stackrel{\eqref{eq:absorption}}{=} (m_* \tensor \idmap)
   \after \name{m} \\ 
   \stackrel{\eqref{eq:absorption}}{=} \name{m \after m^\dag} \\
   = \name{\idmap[C]}.
  \]
  since $m \after m^\dag=\idmap$ by Definition~\ref{def:measurement}.
  The commutativity of the above diagram is then established by the
  following calculation based on the properties of classical structures
  discussed after Definition~\ref{def:classicalstructure}.
  \begin{align*}
    (\epsilon \tensor \idmap) \after (m_* \tensor m) \after
    \eta_X
  & = (\epsilon \tensor \idmap) \after \name{\idmap[C]} \\
  & = (\epsilon \tensor \idmap) \after \delta \after \epsilon^\dag \\
  & = (\idmap \tensor \epsilon) \after \delta \after \epsilon^\dag \\
  & = (\idmap \tensor \epsilon) \after \name{\idmap[C]} \\
  & = (\idmap \tensor \epsilon) \after (m_* \tensor m) \after \eta_X.
  \tag*{$\Box$}
  \end{align*}

\noindent This protocol did not in fact use the choice-of-duals-functor 
on non-compact morphisms, because it only operates on compact parts of
a non-compact object. However, it is entirely feasible that quantum
protocols (or more general constructions in quantum physics)
essentially rely on the choice-of-duals-functor on non-compact parts
when modeled categorically.

\section{Conclusion}
\label{sec:conclusion}

With an eye towards applications in quantum theory, we developed the
notion of a compactly accessible category using the structure of a
factorisation system. It is a category that can
contain objects that are not compact themselves, but are directed
colimits of compact objects, thus allowing for infinite-dimensionality.
Simultaneously, it has a functor that canonically extends the
choice-of-duals on its compact part, and that commutes with a dagger
structure if one is available. The need for such a category was
illustrated by categorically modeling and proving correct a quantum
key distribution protocol. The full structure of dagger compactly
accessible categories was not needed for this specific example, in
particular the extended choice-of-duals functor went unused. But in
general an extended choice-of-duals functor is convenient and even
arguably necessary. Moreover, in the presence of a dagger structure it
hardly puts up more restrictions and hence is essentially for free. 

Several connections to related research present themselves. First, a
compact category has a canonical trace~\cite{abramsky}. Although the
nuclear ideal setting~\cite{abramskyblutepanangaden} seems ideal to
study this phenomenon, perhaps the trace class morphisms can also be
characterized by a colimit property, analogous to the passage from
compact categories to compactly accessible ones. 
Secondly, compactly accessible categories can be seen as a `technical
implementation' of shape theory~\cite{blute}, with the benefit of
actually having concrete structure. One could look for the initial or
terminal such implementation.
Thirdly, the store of qubit-pairs in
Section~\ref{sec:quantumkeydistributioncategorically} strongly
resembles Fock space~\cite{vicary,blutepanangadenseely}, suggesting
that compactly accessible categories might be naturally employed
there. 
Lastly, one could develop a graphical calculus~\cite{selinger}
for (dagger) compactly accessible categories. However, for other
purposes than aiding intuition, this seems a premature optimisation.  

The presented material also indicates some directions for future
research. 
First, a categorical version of the Bell inequalities would lend a
definiteness to the categorical approach to quantum
theory~\cite{abramskycoecke}.  
Secondly, the notion of complete positivity~\cite{selinger} could be
extended to compactly accessible categories. The usual formulation of
a completely positive morphism in a dagger compact category relies
essentially on the category being closed. As a dagger compactly
accessible category is not necessarily closed (\eg $\Cat{Hilb}$), a
different characterization of complete positivity is in order.
Thirdly, the connection to linear logic should be explored. Compact
categories, as special cases of *-autonomous categories, model a large
fragment of linear logic. It is also known that Barr's free
construction of a *-autonomous category provides a model of full
linear logic when one starts with an accessible
category~\cite{barr2}. Thus compactly presentable categories qualify
as likely candidates to model linear logic, perhaps with unusual
properties~\cite{duncan}. 
Fourthly, locally presentable categories are known to be precisely the
models of essentially algebraic theories. Likewise, accessible
categories are precisely the axiomatisations by a basic theory in some
many-sorted first-order logic~\cite{adamekrosicky}. One could look for
similar results that characterise compactly presentable categories and
compactly accessible categories as models of some algebraic theories.
Lastly, extending a compact category to a compactly accessible one
would be a valuable addition to the theory developed in this article,
as we have motivated compactly accessible categories as an extension
of compact ones, but gave only an axiomatic description. A
conceivable starting point 
could be the fact that accessible categories are free cocompletions of
small categories with respect to directed
colimits~\cite[Theorem~2.26]{adamekrosicky}. It would  
involve a completion of a factorisation system with directed
colimits, which would likely involve a study of
dagger factorisation systems in itself, as discussed in
Section~\ref{subsec:structureorproperty}.

\section*{Acknowledgement}

The author is most grateful to Bart Jacobs for his suggestions and
encouraging feedback, and to Michael Barr, Bob Coecke, Jeff Egger and
Isar Stubbe for helping to correct an erroneous example in previous
presentations about this work. The anonymous referees made some
detailed suggestions which improved the paper appreciably.
Finally Rick Blute provided some helpful comments.

\bibliographystyle{alpha}
\bibliography{cptacccats-lmcs2}

\vspace{-40 pt}
\end{document}